\documentclass[onecolumn]{aastex61}
\usepackage{amsmath}

\shorttitle{Proper Motions and the Matter Power Spectrum}
\shortauthors{Darling \& Truebenbach}
\begin{document}
\title{All Transverse Motion is Peculiar:  Connecting the Proper Motions of Galaxies to the Matter Power Spectrum}

\correspondingauthor{Jeremy Darling}
\email{jeremy.darling@colorado.edu}

\author[0000-0003-2511-2060]{Jeremy Darling}
\affil{Center for Astrophysics and Space Astronomy \\
Department of Astrophysical and Planetary Sciences \\
University of Colorado, 389 UCB \\
Boulder, CO 80309-0389, USA}
\author{Alexandra E. Truebenbach}
\affil{Center for Astrophysics and Space Astronomy \\
Department of Astrophysical and Planetary Sciences \\
University of Colorado, 389 UCB \\
Boulder, CO 80309-0389, USA}

\begin{abstract}
In an isotropic and homogeneous Hubble expansion, all transverse motion is peculiar.  Like the radial peculiar velocities of galaxies, 
transverse peculiar velocities are a means to trace the density of matter that does not rely on light tracing mass.  
Unlike radial peculiar velocity measurements that require precise redshift-independent distances in order to distinguish between 
the Hubble expansion and the observed redshift, transverse peculiar velocities can be measured using redshifts alone as a proxy for distance.  
Extragalactic proper motions can therefore directly measure peculiar velocities and probe the matter
power spectrum.  Here we develop two-point transverse velocity correlation statistics and demonstrate their dependence on the 
matter power spectrum.  We predict the power in these correlation statistics as a function of the physical separation, angular separation, and distance of pairs of 
galaxies and demonstrate that the effect of large scale structure on transverse motions is best measured for pairs of objects with 
comoving separations less than about 50 Mpc.  Transverse peculiar velocities induced by large scale structure
should be observable as proper motions using long baseline radio interferometry or space-based optical astrometry.  
\end{abstract}

\keywords{astrometry --- cosmology:  observations --- cosmology:  theory
--- large-scale structure of universe --- proper motions --- techniques: high angular resolution }

\section{Introduction}

In a pure isotropic and homogeneous Hubble expansion, all apparent motion is radial.  
Therefore, proper motions of galaxies are the observable manifestation of peculiar velocities transverse to the line of sight
induced by the density inhomogeneities of large scale structure.  Extragalactic proper motions can be induced by a moving
observer, an accelerating observer, or gravitational waves, which imprint a large-scale (dipole or quadrupole) proper motion
signal on the sky \citep{ding2009,darling2013,truebenbach2017,darling2018}.  In contrast, 
the proper motion correlation of pairs of galaxies will imprint on all angular scales with 
a predictable dependence on pair separation that can be distinguished from these observer-induced signatures.  

Studies of radial peculiar velocities require precise measurements of distances in order to disentangle the cosmological 
recession velocity from any peculiar velocity \citep[e.g.,][]{strauss1995,dekel1997}.  Studies of transverse peculiar velocities, on the other hand, do not rely on accurate 
distance measurements and therefore, in principle, avoid a significant systematic error and can be made independent 
of cosmological model or extragalactic distance ladder.  That said, proper motions are difficult to measure, particularly at the level that peculiar velocities are expected to arise:   $v_{\rm pec} \sim 300$ km s$^{-1}$ is 
a proper motion of $\mu \sim 60$~$\mu$as~yr$^{-1}$ at 1~Mpc ($\mu = v_{\rm pec}/D_M$, neglecting 
the rate of change of the proper motion distance $D_M$).  The distance at which transverse peculiar velocities 
could conceivably be measured in large astrometric surveys is about 100 Mpc, which is similar to radial peculiar velocity studies
\citep[e.g.,][]{tully2014}.  Measurements at much larger distances may be possible using gravitational lensing 
\citep[e.g.][]{kochanek1996,mediavilla2015,mediavilla2016}.

Peculiar velocities relate directly to matter:  it is gravity (dominated by dark matter) that drives peculiar motions.  Unlike
the two-point spatial correlation function, which depends strongly on small spatial scales, the peculiar velocity correlation 
functions get most of their power from larger spatial scales and therefore are better related to the linear density
perturbation growth regime and more easily connected to the matter power spectrum \citep[e.g.,][]{dodelson2003}.

Here we derive observable correlation statistics between pairs of extragalactic proper motions and relate these statistics to the 
matter power spectrum (Sections \ref{sec:correlation} and \ref{sec:results}).  We explore observational strategies for detecting the peculiar velocities of galaxies induced by large scale structure in Section \ref{sec:discussion}.  
We parameterize the Hubble expansion today as $H_0 = 100$~$h$~km~s$^{-1}$~Mpc$^{-1}$
and assume a geometrically flat universe with $\Omega_m = 0.3$ and $\Omega_\Lambda = 0.7$.

\section{A Transverse Peculiar Velocity Correlation Function}\label{sec:correlation}

Proper motions measure the transverse peculiar velocity, after observer-induced effects, such as the secular 
aberration drift and extragalactic parallax, and cosmological effects, such as primordial gravitational waves, 
are removed \citep{ding2009,darling2013,darling2014,bower2015,truebenbach2017,darling2018}.  But transverse velocities for 
different objects are not coplanar, so two-point correlations of pairs
of transverse velocities will mix transverse and radial peculiar velocity correlation functions.  This mixing of vector
components in a spherical geometry is one reason why even 1D line-of-sight peculiar velocity studies are able
to produce reasonable density maps \citep[e.g.,][]{dekel1997}.

We define a two-point correlation statistic $\xi_{v,\perp}$ that projects the transverse velocity of each object $\vec{v}_\perp(\vec{x}_i)$  
onto the space vector connecting the two objects, $\vec{x} = \vec{x}_1 - \vec{x}_2$, as
\begin{equation}  \label{eqn:statistic} 
  \xi_{v,\perp}(\vec{x}_1,\vec{x}_2) = \langle (\vec{v}_\perp(\vec{x}_1) \cdot \hat{x}) (\vec{v}_\perp(\vec{x}_2) \cdot \hat{x}) \rangle,
\end{equation}
where the brackets indicate an average over all pairs with separation $|\vec x|$. 
This statistic will produce negative values for pairs of objects that are converging or diverging along $\hat{x}$ 
and positive values for co-streaming motions.  Unlike the radial peculiar velocity correlation function that 
depends directly on observable Doppler shifts $v(\vec{x}_i)$, $\xi_v(\vec{x}_1,\vec{x}_2) = \langle v(\vec{x}_1)\, v(\vec{x}_2) \rangle$, 
two-dimensional transverse velocities must be reduced in dimensionality, and the choice for this reduction is 
non-unique.  Since transverse velocities (observed as proper motions) of pairs of objects separated on the sky 
are not co-planar and are induced by the density inhomogeneities of large scale structure, we project 
the transverse velocities onto the space unit vector that connects the two objects, $\hat{x}$.    Another option is 
to simply take a dot product of the two transverse velocity vectors, 
$\xi^\prime_{v,\perp}(\vec{x}_1,\vec{x}_2) \equiv \langle \vec{v}_\perp(\vec{x}_1) \cdot \vec{v}_\perp(\vec{x}_2) \rangle$
 (see Section \ref{sec:alt_stat} for the derivation and results of this approach).  

In what follows we connect this correlation statistic to the matter power spectrum in order
to predict the correlated signals that should be observed in extragalactic proper motions and to 
connect observations to the matter power spectrum in a distance ladder-independent way.  
In Section \ref{subsec:derivation} we show that, in the linear regime of structure growth, which is 
a fair assumption for peculiar velocity statistics, Equation \ref{eqn:statistic} can be reduced to two 
wavenumber integrals of the matter power spectrum and the derivatives of spherical Bessel functions:
\begin{eqnarray}
  \xi_{v,\perp}(\vec{x}_1,\vec{x}_2) 
      = - f^2 H_0^2 \left[\sin^2\theta_1\sin^2\theta_2\ \int_0^\infty {dk\over 2\pi^2k}\ P(k)\ k\ j_0^{\prime\prime}(kx)\right. \nonumber \\
              +\left.\frac{1}{4}\sin2\theta_1\sin2\theta_2 \ \int_0^\infty {dk\over 2\pi^2k}\ P(k)\ {j_0^\prime(kx)\over x}\right] .
\end{eqnarray}
Terms are defined below.

\subsection{Derivation}\label{subsec:derivation}

The following derivation is an adaptation of and closely follows the treatment of radial peculiar 
velocities by \citet{dodelson2003} and relies on two assumptions:  (1) linear growth of density perturbations 
$\delta = \delta \rho/\rho$, which can be directly related to the Fourier components (spatial frequencies $\vec k$) of 
peculiar velocity at low redshift,
\begin{equation}\label{eqn:linear}
  \vec{v}(\vec{k}) = i f  H_0\ \delta(\vec{k}) {\vec{k} \over k^2}, 
\end{equation}
where $f$ is the dimensionless linear growth rate approximated by $f = \Omega^{0.6}_m$,
and (2) low redshift, $z\lesssim1$.  

\begin{figure}
\epsscale{0.9}
\plotone{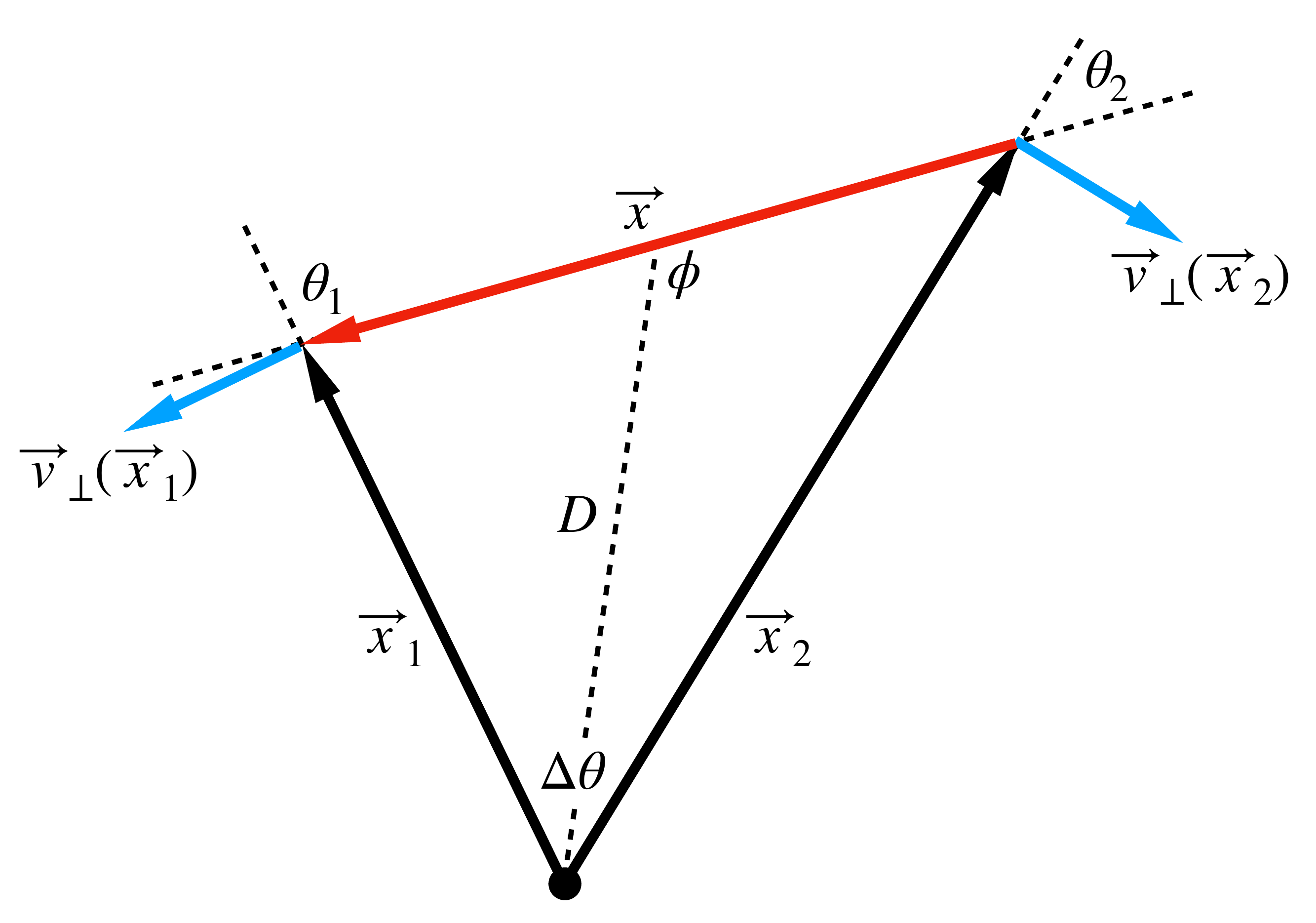}
\caption{The vectors and angles associated with a pair of objects with transverse velocities.
The velocities have components out of the plane containing the two objects and the origin and are not necessarily coplanar.  
$D$ is the distance to the pair midpoint, $\phi$ is the pair vector orientation angle, and $\Delta\theta = \theta_1 - \theta_2$.  
}\label{fig:diagram}
\end{figure}
 
Any velocity vector can be decomposed into radial and transverse components:
\begin{equation}
  \vec{v}(\vec x_i) = \vec{v}_\perp(\vec x_i) + \vec{v}_\parallel(\vec x_i) = \vec{v}_\perp(\vec x_i) + (\vec{v}(\vec x_i)\cdot\hat{x}_i)\,\hat{x}_i
\end{equation}
where $\hat{x}_i$ is the (radial) unit space vector to the object $i$ with velocity $\vec{v}(\vec x_i)$.  The transverse 
velocity vector can therefore be expressed in terms of the total velocity vector and the direction to the object:
\begin{equation}
  \vec{v}_\perp(\vec x_i)  =  \vec{v}(\vec x_i) - (\vec{v}(\vec x_i)\cdot\hat{x}_i)\, \hat{x}_i.  \label{eqn:vperp}
\end{equation}

Inserting Equation \ref{eqn:vperp} into the correlation statistic (Equation \ref{eqn:statistic}), we find that
\begin{equation}
  \xi_{v,\perp}(\vec{x}_1,\vec{x}_2) = \langle \left(\left[\vec{v}(\vec x_1) - (\vec{v}(\vec x_1)\cdot\hat{x}_1) \hat{x}_1\right]\cdot\hat{x}\right)
                                                                \left(\left[\vec{v}(\vec x_2) - (\vec{v}(\vec x_2)\cdot\hat{x}_2)\hat{x}_2 \right]\cdot\hat{x} \right) \rangle.
\end{equation}
Defining angles $\theta_1$ and $\theta_2$ in terms of the projections of $\hat{x}_1$ and $\hat{x}_2$ onto $\hat{x}$ (Figure \ref{fig:diagram}), 
$\cos\theta_1 \equiv - \hat{x}_1 \cdot \hat{x}$ and $\cos\theta_2 \equiv - \hat{x}_2 \cdot \hat{x}$,\footnote{These slightly unusual 
definitions arise from the definition $\vec{x} = \vec{x}_1 - \vec{x}_2$, the dot product 
$\hat{x}_i\cdot\hat{x} = \cos(\pi-\theta_i) = -\cos\theta_i$ and the requirement that the observed angle between the two 
objects be $\Delta\theta = \theta_1 - \theta_2$. See Figure \ref{fig:diagram}.} and expanding the dot products, 
\begin{equation}
  \xi_{v,\perp}(\vec{x}_1,\vec{x}_2) = \langle \left(\vec{v}(\vec x_1)\cdot\hat{x} + \vec{v}(\vec x_1)\cdot\hat{x}_1\cos\theta_1\right)
                                                                \left(\vec{v}(\vec x_2)\cdot\hat{x} + \vec{v}(\vec x_2)\cdot\hat{x}_2\cos\theta_2 \right) \rangle.  
\end{equation}
This expression expands into four parts that can be treated separately:
\begin{eqnarray}\label{eqn:allparts}
  \xi_{v,\perp}(\vec{x}_1,\vec{x}_2) = \langle \overbrace{(\vec{v}(\vec x_1)\cdot\hat{x}) (\vec{v}(\vec x_2)\cdot\hat{x})}^{(i)} \rangle 
                                        + \langle \overbrace{(\vec{v}(\vec x_1)\cdot\hat{x})(\vec{v}(\vec x_2)\cdot\hat{x}_2\cos\theta_2)}^{(ii)} \rangle 
                                        + \langle \overbrace{(\vec{v}(\vec x_2)\cdot\hat{x})(\vec{v}(\vec x_1)\cdot\hat{x}_1\cos\theta_1)}^{(iii)} \rangle \nonumber \\
                                        + \langle \overbrace{(\vec{v}(\vec x_1)\cdot\hat{x}_1)(\vec{v}(\vec x_2)\cdot\hat{x}_2)\cos\theta_1\cos\theta_2}^{(iv)}
   \rangle.  
\end{eqnarray}
For each part of this expression (following \citet{dodelson2003}), we recast the velocity vectors in terms of their Fourier components, 
\begin{equation}\label{eqn:fourier}
  \vec{v}(\vec{x}_i) = \int {d^3k\over(2\pi)^3}\ e^{i \vec{k}\cdot\vec{x}_i}\ \vec{v}(\vec{k}), 
\end{equation}
and employ linear theory for $\vec{v}(\vec{k})$ to relate it to the density perturbation $\delta(\vec k)$ (Equation \ref{eqn:linear}). 

\subsubsection{Part $(i)$}  
Part $(i)$ is the transverse counterpart to the radial peculiar velocity two point correlation function, 
  $\xi_v(\vec{x}_1,\vec{x}_2)= \langle (\vec{v}(\vec x_1)\cdot\hat{x}_1)\,(\vec{v}(\vec x_2)\cdot\hat{x}_2) \rangle$.  Rather than 
projecting the 3D velocities onto the radial coordinate, $(i)$ projects the 3D velocities onto the vector connecting the 
two objects, $\hat x$.  For pairs of objects with small angular separations, $\hat x$ be nearly coplanar with the observed 
proper motion vectors.  

Using the Fourier transforms of the velocity vectors (Eqn.\ \ref{eqn:fourier}) followed by the linear velocity-density relation 
(Eqn.\ \ref{eqn:linear}), part $(i)$ of the correlation statistic becomes
\begin{eqnarray}
  \langle (\vec{v}(\vec x_1)\cdot\hat{x}) (\vec{v}(\vec x_2)\cdot\hat{x}) \rangle 
      &=& \int {d^3k\over(2\pi)^3}\ e^{i \vec{k}\cdot\vec{x}_1} 
         \int {d^3k^\prime\over(2\pi)^3}\ e^{-i \vec{k}^\prime\cdot\vec{x}_2}\ 
         \langle (\vec{v}(\vec{k}) \cdot \hat{x}) (\vec{v}^{\,*}(\vec{k}^\prime) \cdot \hat{x}) \rangle \\
      &=& f^2 H_0^2\ \int {d^3k\over(2\pi)^3}\ e^{i \vec{k}\cdot\vec{x}_1} 
         \int {d^3k^\prime\over(2\pi)^3}\ e^{-i \vec{k}^\prime\cdot\vec{x}_2}\ 
         \langle \delta(\vec{k}) \delta^*(\vec{k}^\prime) \rangle {(\vec{k} \cdot \hat{x})(\vec{k}^\prime\cdot\hat{x})\over 
               k^2 k^{\prime\,2}}.
\end{eqnarray}
Using what is often the definition of the matter power spectrum, $P(k)$, in terms of the density fluctuation variance
and the Dirac delta function $\delta^3(\vec{k}-\vec{k}^\prime)$, 
\begin{equation}
  \langle \delta(\vec{k}) \delta^*(\vec{k}^\prime) \rangle = (2\pi)^3 \delta^3(\vec{k}-\vec{k}^\prime) P(k), 
\end{equation}
we can reduce part $(i)$ to integrals in a single wavenumber:
\begin{equation}
  \langle (\vec{v}(\vec x_1)\cdot\hat{x}) (\vec{v}(\vec x_2)\cdot\hat{x}) \rangle 
      = f^2 H_0^2\ \int_0^\infty {dk\, k^2 \over(2\pi)^3}\ P(k) 
         \int d\Omega_k\ e^{i \vec{k}\cdot\vec{x}}\  {(\vec{k} \cdot \hat{x})^2\over k^4}.
\end{equation}
Since the wavenumber can be written
\begin{equation}
  \vec{k} = e^{-i\vec{k}\cdot\vec{x}}\  {1\over i}\, \frac{\partial}{\partial \vec{x}}\, e^{i\vec{k}\cdot\vec{x}}\ ,
\end{equation}
using index notation and the Einstein convention, we have
\begin{equation}
  e^{i \vec{k}\cdot\vec{x}}\  (\vec{k} \cdot \hat{x}) (\vec{k} \cdot \hat{x}) = 
         -\hat{x}_i \hat{x}_j \frac{\partial^2}{\partial x_i \partial x_j} e^{i \vec{k}\cdot\vec{x}}
\end{equation}
and part $(i)$ becomes
\begin{equation}
    \langle (\vec{v}(\vec x_1)\cdot\hat{x}) (\vec{v}(\vec x_2)\cdot\hat{x}) \rangle 
      = - f^2 H_0^2\ \int_0^\infty {dk\over (2\pi)^3 k^2}\ P(k)\ 
           \hat{x}_i \hat{x}_j \frac{\partial^2}{\partial x_i \partial x_j} \int d\Omega_k\ e^{i \vec{k}\cdot\vec{x}}.
\end{equation}
At this stage, the only difference between  $\xi_v(\vec{x}_1,\vec{x}_2)$ and the above are the unit vectors:
for the radial peculiar velocity correlation function, these are the radial unit vectors $\hat{x}_1$ and $\hat{x}_2$ \citep{dodelson2003},
while for part $(i)$ of the transverse peculiar velocity correlation function, it is the unit vector $\hat{x}$.  
The angular integral is 
\begin{equation}
  \int d\Omega_k\ e^{i \vec{k}\cdot\vec{x}} = \int_0^{2\pi} d\phi \int_{-1}^{+1} d\mu\ e^{i k x \mu} 
                                                             = 4\pi j_0(kx)
\end{equation}
where $j_0(kx) = \sin(kx)/kx$ is the $\ell=0$ spherical Bessel function.  The derivatives with respect to the components 
of $\vec x$ can be rewritten in terms of derivatives of the argument of $j_0$, $kx$:
\begin{equation}
  \frac{\partial^2}{\partial x_i \partial x_j} j_0(kx) = k^2\left(\left[\delta_{ij}-\hat{x}_i\hat{x}_j\right] {j_0^\prime(kx)\over kx}
                                                                                               +\hat{x}_i\hat{x}_j j_0^{\prime\prime}(kx)\right).
\end{equation}
Thus, 
\begin{equation}
    \langle (\vec{v}(\vec x_1)\cdot\hat{x}) (\vec{v}(\vec x_2)\cdot\hat{x}) \rangle 
      = - f^2 H_0^2\ \int_0^\infty {dk\over 2\pi^2}\ P(k)\ 
           \hat{x}_i \hat{x}_j \left(\left[\delta_{ij}-\hat{x}_i\hat{x}_j\right] {j_0^\prime(kx)\over kx}
                                                                                               +\hat{x}_i\hat{x}_j j_0^{\prime\prime}(kx)\right) 
\end{equation}
Summing over $i$ and $j$ causes the $j_0^\prime(kx)$ term to vanish.  We will label this result for part $(i)$ $\xi_{v,(i)}$:
\begin{equation}
      \xi_{v,(i)}(x) = - f^2 H_0^2\ \int_0^\infty {dk\over 2\pi^2k}\ P(k)\ k\ j_0^{\prime\prime}(kx).
\end{equation}
It is instructive to examine the behavior of $P(k)$, $j_0^{\prime\prime}(kx)$, and 
the integral of their product, which we will do below (Section \ref{subsec:sum}).

\subsubsection{Part $(ii)$}  
For part $(ii)$, we have
\begin{eqnarray}
  \langle(\vec{v}(\vec x_1)\cdot\hat{x})(\vec{v}(\vec x_2)\cdot\hat{x}_2\cos\theta_2)\rangle
     &=& \int {d^3k\over(2\pi)^3}\ e^{i \vec{k}\cdot\vec{x}_1} 
         \int {d^3k^\prime\over(2\pi)^3}\ e^{-i \vec{k}^\prime\cdot\vec{x}_2}\ 
         \langle (\vec{v}(\vec{k}) \cdot \hat{x}) (\vec{v}^{\,*}(\vec{k}^\prime) \cdot \hat{x}_2\cos\theta_2) \rangle \\
      &=& f^2 H_0^2\ \int {d^3k\over(2\pi)^3}\ e^{i \vec{k}\cdot\vec{x}_1} 
         \int {d^3k^\prime\over(2\pi)^3}\ e^{-i \vec{k}^\prime\cdot\vec{x}_2}\ 
         \langle \delta(\vec{k}) \delta^*(\vec{k}^\prime) \rangle {(\vec{k} \cdot \hat{x})(\vec{k}^\prime\cdot\hat{x}_2)\cos\theta_2\over 
               k^2 k^{\prime\,2}}\\
      &=& f^2 H_0^2\ \int_0^\infty {dk\, k^2 \over(2\pi)^3}\ P(k) 
         \int d\Omega_k\ e^{i \vec{k}\cdot\vec{x}}\  {(\vec{k} \cdot \hat{x}) (\vec{k} \cdot \hat{x}_2)\cos\theta_2\over k^4} \\
      &=& - f^2 H_0^2\ \int_0^\infty {dk\over (2\pi)^3 k^2}\ P(k)\ 
           \hat{x}_i \hat{x}_{2,j} \cos\theta_2\ \frac{\partial^2}{\partial x_i \partial x_j} \int d\Omega_k\ e^{i \vec{k}\cdot\vec{x}} \\
      &=& - f^2 H_0^2\ \int_0^\infty {dk\over 2\pi^2}\ P(k)\ 
           \hat{x}_i \hat{x}_{2,j} \cos\theta_2\ \left(\left[\delta_{ij}-\hat{x}_i\hat{x}_j\right] {j_0^\prime(kx)\over kx}
                                                                                               +\hat{x}_i\hat{x}_j j_0^{\prime\prime}(kx)\right) \\
      &=& - f^2 H_0^2\ \int_0^\infty {dk\over 2\pi^2}\ P(k)\ (\hat{x}_2\cdot\hat{x}) \cos\theta_2\  j_0^{\prime\prime}(kx).
\end{eqnarray}
Since $\hat{x}_2\cdot\hat{x} = -\cos\theta_2$, we obtain an expression for part $(ii)$ that is very similar to the result
for part $(i)$ modulo a factor of $-\cos^2\theta_2$:
\begin{eqnarray}
   \langle(\vec{v}(\vec x_1)\cdot\hat{x})(\vec{v}(\vec x_2)\cdot\hat{x}_2\cos\theta_2)\rangle 
       &=& +\cos^2\theta_2\ f^2 H_0^2\ \int_0^\infty {dk\over 2\pi^2k}\ P(k)\ k\  j_0^{\prime\prime}(kx)\\
       &=& -\cos^2\theta_2\ \xi_{v,(i)}(x).
\end{eqnarray}

\subsubsection{Part $(iii)$}   
Part $(iii)$ has a nearly identical result to part $(ii)$:
\begin{equation}
  \langle(\vec{v}(\vec x_2)\cdot\hat{x})(\vec{v}(\vec x_1)\cdot\hat{x}_1\cos\theta_1)\rangle 
      = -\cos^2\theta_1\ \xi_{v,(i)}(x).
\end{equation}

\subsubsection{Part $(iv)$}    
Part $(iv)$ follows the same procedure until the stage where we sum over $i$ and $j$:
\begin{eqnarray}
 \langle(\vec{v}(\vec x_1)\cdot\hat{x}_1)(&\vec{v}(\vec x_2)&\cdot\hat{x}_2) \cos\theta_1 \cos\theta_2\rangle \nonumber\\ 
      &=& - f^2 H_0^2\int_0^\infty {dk\over 2\pi^2}\ P(k)\,
      \hat{x}_{1,i}\hat{x}_{2,j}\cos\theta_1\cos\theta_2\left(\left[\delta_{ij}-\hat{x}_i\hat{x}_j\right] {j_0^\prime(kx)\over kx}
                                                                                               +\hat{x}_i\hat{x}_j j_0^{\prime\prime}(kx)\right) \\
      &=& - f^2 H_0^2\int_0^\infty {dk\over 2\pi^2}\ P(k)\,
      \cos\theta_1\cos\theta_2\left(\left[\cos(\theta_1-\theta_2)-\cos\theta_1\cos\theta_2\right] {j_0^\prime(kx)\over kx} 
          +\cos\theta_1\cos\theta_2 j_0^{\prime\prime}(kx)\right) 
          \label{eqn:part_iv_3}\\
      &=& - f^2 H_0^2\int_0^\infty {dk\over 2\pi^2 k}\ P(k)\,
                 \left(\sin\theta_1\sin\theta_2 \cos\theta_1\cos\theta_2\ {j_0^\prime(kx)\over x}
                                                                                               +\cos^2\theta_1\cos^2\theta_2\ k\  j_0^{\prime\prime}(kx)\right),
\end{eqnarray}
where $\cos(\theta_1-\theta_2) = \hat{x}_1\cdot\hat{x}_2$, which comes from the inner angles of the 
$\vec{x}_1$-$\vec{x}_2$-$\vec{x}$ triangle, is invoked in Equation \ref{eqn:part_iv_3}.  We define 
$\xi_{v,(iv)}(x)$ to be the integral involving $j_0^\prime(kx)$ to obtain an expression for part $(iv)$:
\begin{equation}
 \langle(\vec{v}(\vec x_1)\cdot\hat{x}_1)(\vec{v}(\vec x_2)\cdot\hat{x}_2) \cos\theta_1 \cos\theta_2\rangle 
    = \sin\theta_1\sin\theta_2 \cos\theta_1\cos\theta_2\ \xi_{v,(iv)}(x) + \cos^2\theta_1\cos^2\theta_2\ \xi_{v,(i)}(x).
\end{equation}

\begin{figure}
\epsscale{0.575}
\plotone{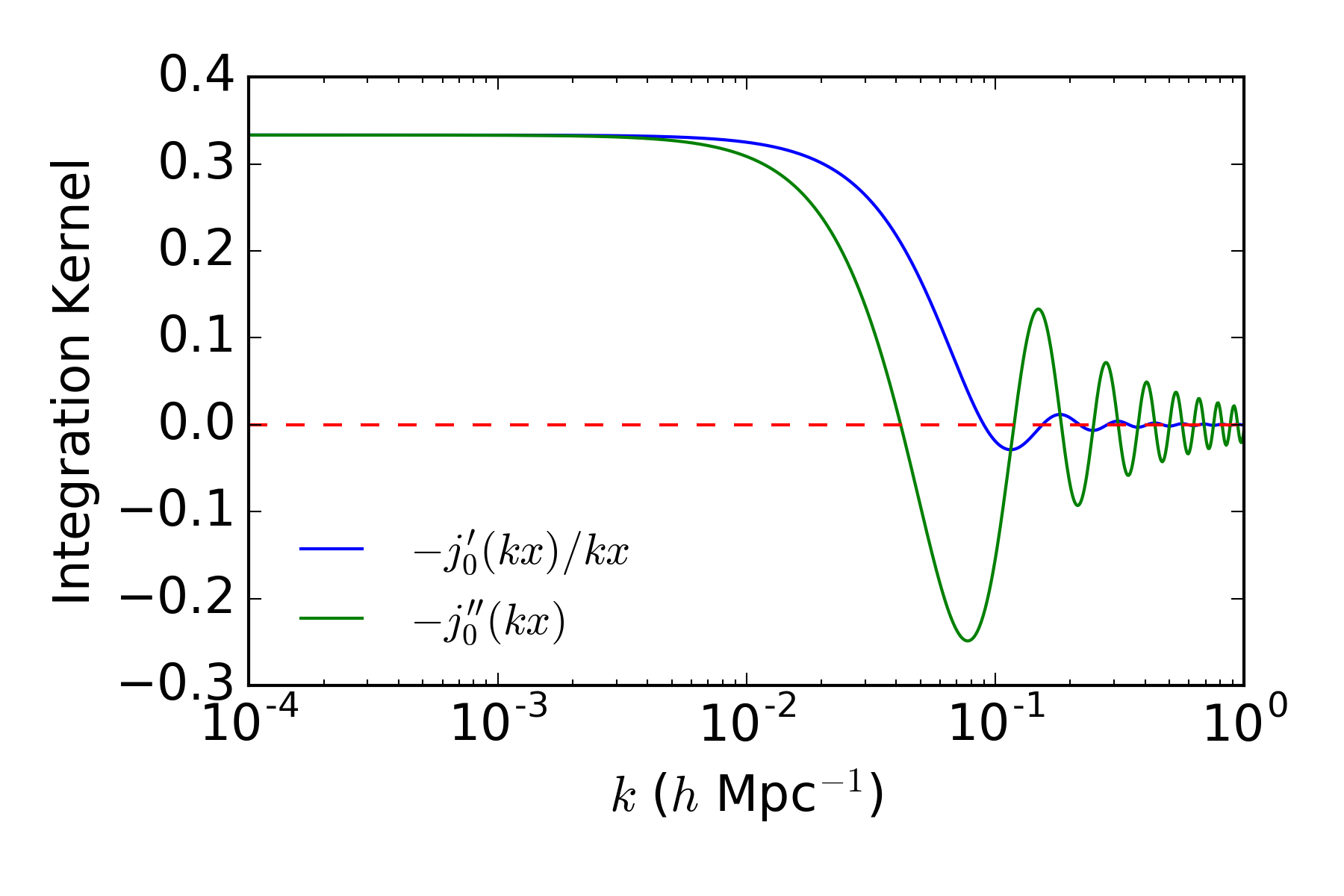}
\plotone{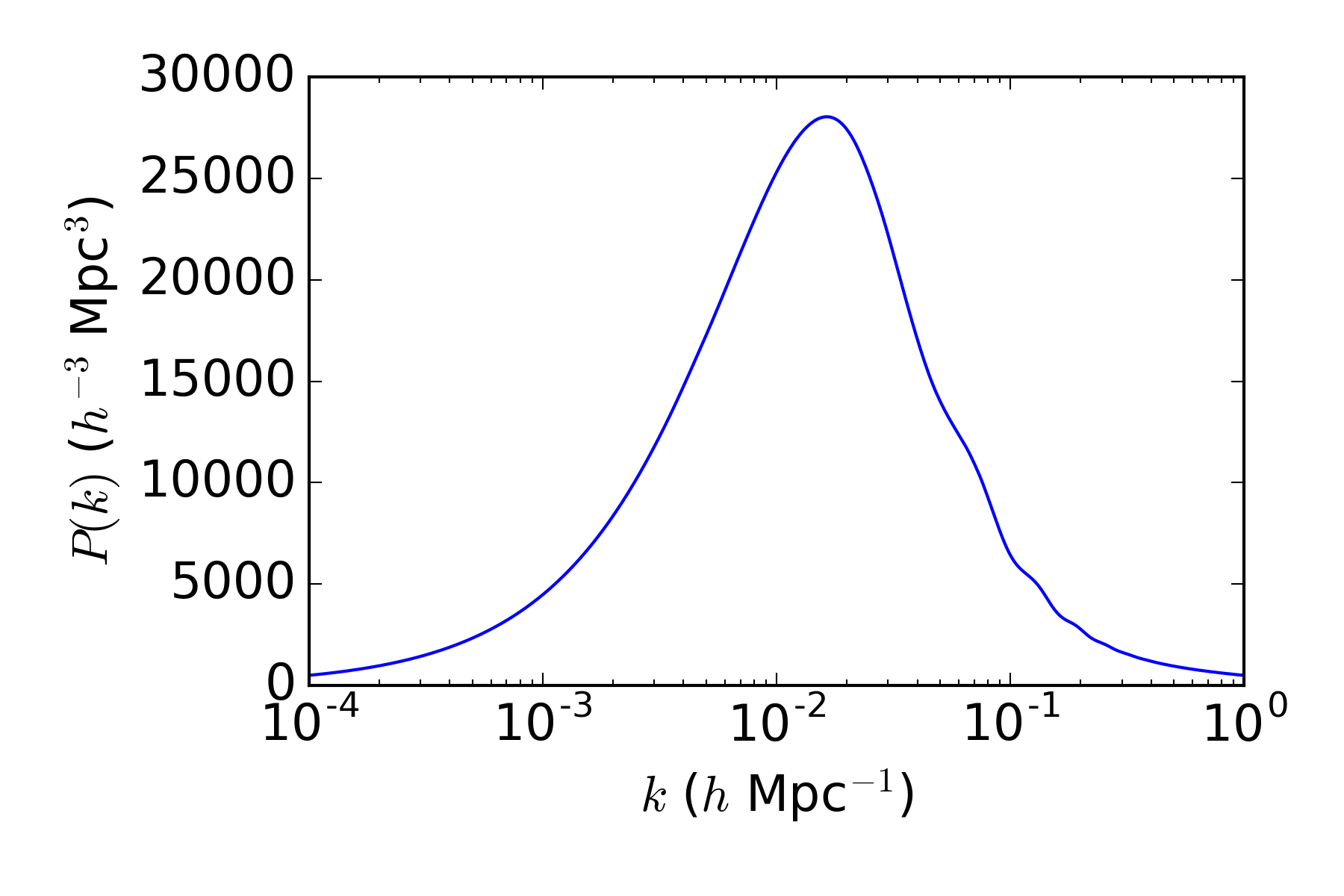}
\plotone{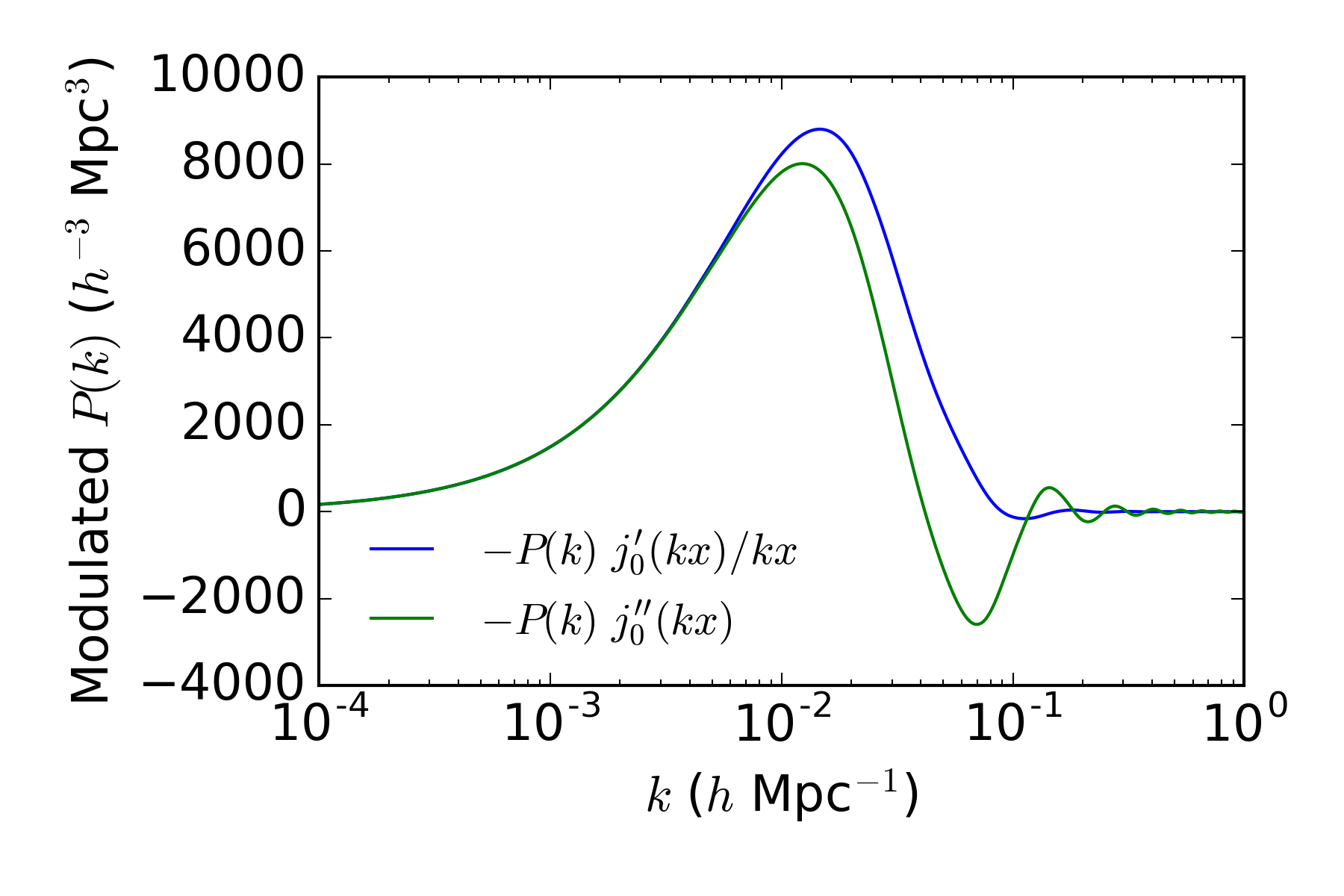}
\plotone{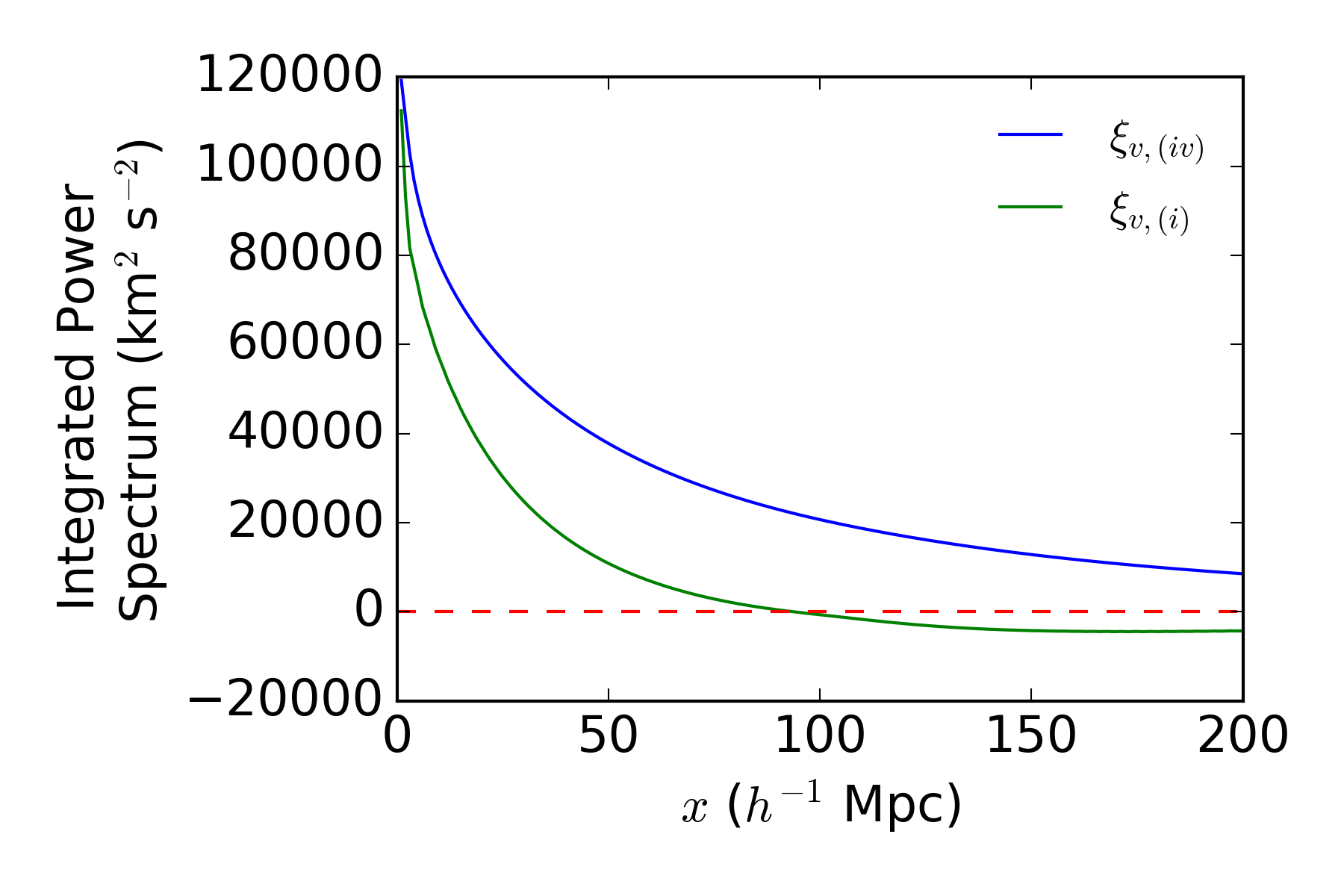}
\caption{Top Left: Integration kernels for the transverse peculiar velocity correlation function (Equation \ref{eqn:result2}) 
versus wavenumber for objects separated by $x = 50$~$h^{-1}$~Mpc.
Top Right:   Matter power spectrum $P(k)$ versus wavenumber obtained from CAMB (Section \ref{subsec:sum}).  
Bottom Left:  Matter power spectrum modulated by the integration kernels for $x = 50$~$h^{-1}$~Mpc.
Bottom Right:  Integrated modulated power spectrum (the two terms in Equation \ref{eqn:result}, omitting the prefactors on
$\xi_{v,(i)}$ and  $\xi_{v,(iv)}$) versus pair separation $x$.  
}\label{fig:powerspec}
\end{figure}

\subsubsection{Sum of Parts}   \label{subsec:sum}
Summing all parts of Equation \ref{eqn:allparts}, we have
\begin{equation}
  \xi_{v,\perp}(\vec{x}_1,\vec{x}_2) 
      = \sin^2\theta_1\sin^2\theta_2\ \xi_{v,(i)}(x) 
              +  \frac{1}{4}\sin2\theta_1\sin2\theta_2 \ \xi_{v,(iv)}(x) \label{eqn:result}
\end{equation}
or
\begin{eqnarray}
  \xi_{v,\perp}(\vec{x}_1,\vec{x}_2) 
      = - f^2 H_0^2 \left[\sin^2\theta_1\sin^2\theta_2\ \int_0^\infty {dk\over 2\pi^2k}\ P(k)\ k\ j_0^{\prime\prime}(kx)\right. \nonumber \\
              +\left.\frac{1}{4}\sin2\theta_1\sin2\theta_2 \ \int_0^\infty {dk\over 2\pi^2k}\ P(k)\ {j_0^\prime(kx)\over x}\right] . \label{eqn:result2}
\end{eqnarray}
The $\xi_{v,(i)}(x)$ term represents the motion along the pair separation axis $\hat x$, and the 
$\xi_{v,(iv)}(x)$ term represents the motion perpendicular to $\hat x$ (defining a plane).  
It is instructive to examine the two kernels in this expression to see where most of the power lies in 
the correlation of transverse peculiar velocities.  Figure \ref{fig:powerspec} (top left) shows the integration kernels for the two 
terms above, assuming $x = 50$~$h^{-1}$~Mpc.  While power will be included on all scales, there is a suppression of power on spatial 
frequencies $k \gtrsim 0.1$~$h$~Mpc$^{-1}$.  As with the radial velocity correlation function, the transverse velocity 
correlation is less sensitive to the nonlinear density perturbation regime than the density correlation function.   
 
Figure \ref{fig:powerspec} (top right) shows the matter power spectrum $P(k)$ obtained from 
the online Code for Anisotropies in the Microwave Background 
(CAMB)\footnote{Lewis, A. \& Challinor, A., August 2017 version.} for $z=0$, $H_0 = 70$~km~s$^{-1}$~Mpc$^{-1}$, 
$\Omega_b = 0.0462$,  $\Omega_{CDM} = 0.2538$, and  $\Omega_\Lambda = 0.7$.  
Figure \ref{fig:powerspec}  (bottom left)
shows the kernel-weighted power spectrum for each term $\xi_{v,(i)}$ and $\xi_{v,(iv)}$ versus wavenumber $k$.
Figure \ref{fig:powerspec} (bottom right) shows the wavenumber-integrated
kernel-weighted  power spectrum for each term $\xi_{v,(i)}$ and $\xi_{v,(iv)}$ in Equation \ref{eqn:result} 
(omitting prefactors, which depend on individual pairs) versus physical separation $x$.  It should be stressed that 
while these terms in the transverse peculiar velocity correlation function peak at the smallest pair separations, the scales driving the 
correlated motions are dominated by much larger scale structure, $k\lesssim0.1$~$h$~Mpc$^{-1}$.  
For example, a fairly close pair of galaxies will show highly correlated
peculiar motion as they both respond to the density enhancement of the local filament or supercluster.

\section{Results}\label{sec:results}

\subsection{Equidistant Pairs}\label{subsec:equidistant}

\begin{figure}
\epsscale{0.575}
\plotone{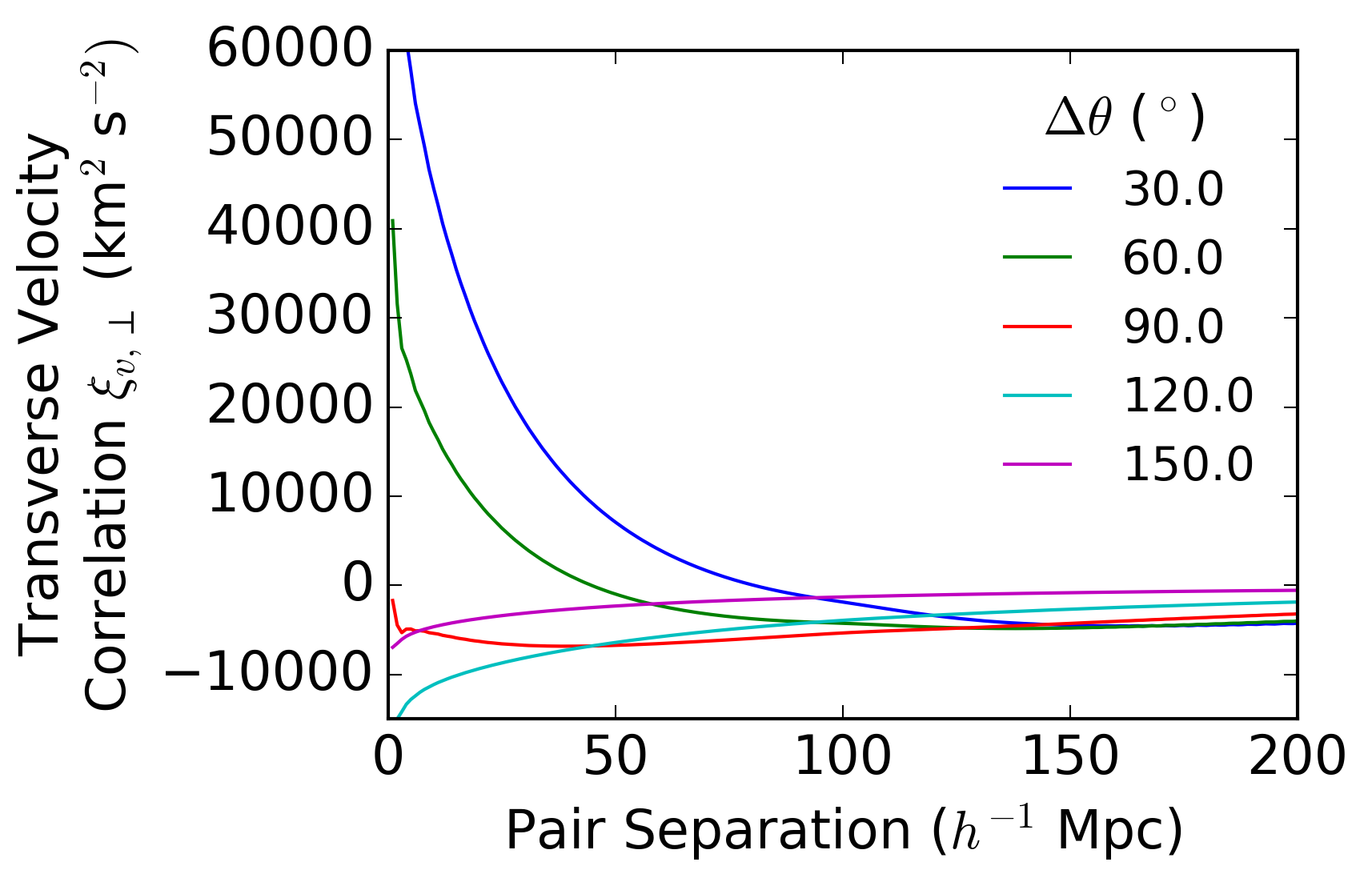}
\plotone{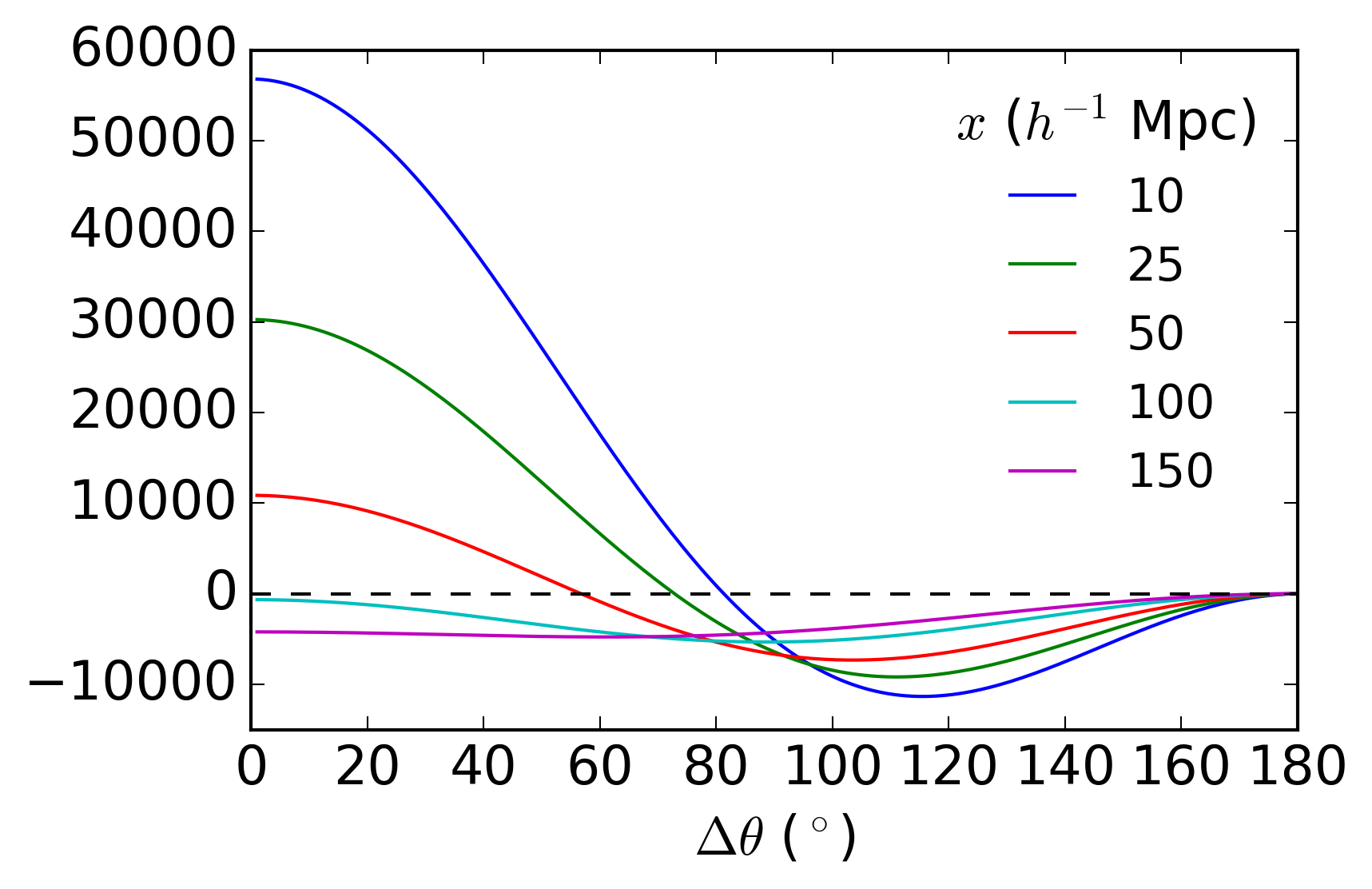}
\caption{Transverse velocity correlation $\xi_{v,\perp}(\vec{x}_1,\vec{x}_2)$ versus physical separation $x$  (left) and versus angular separation (right) for pairs of objects at equal distance ($|\vec{x}_1| = |\vec{x}_2|$).
}\label{fig:xi_equal}
\end{figure}

\begin{figure}
\includegraphics[width=0.44\textwidth]{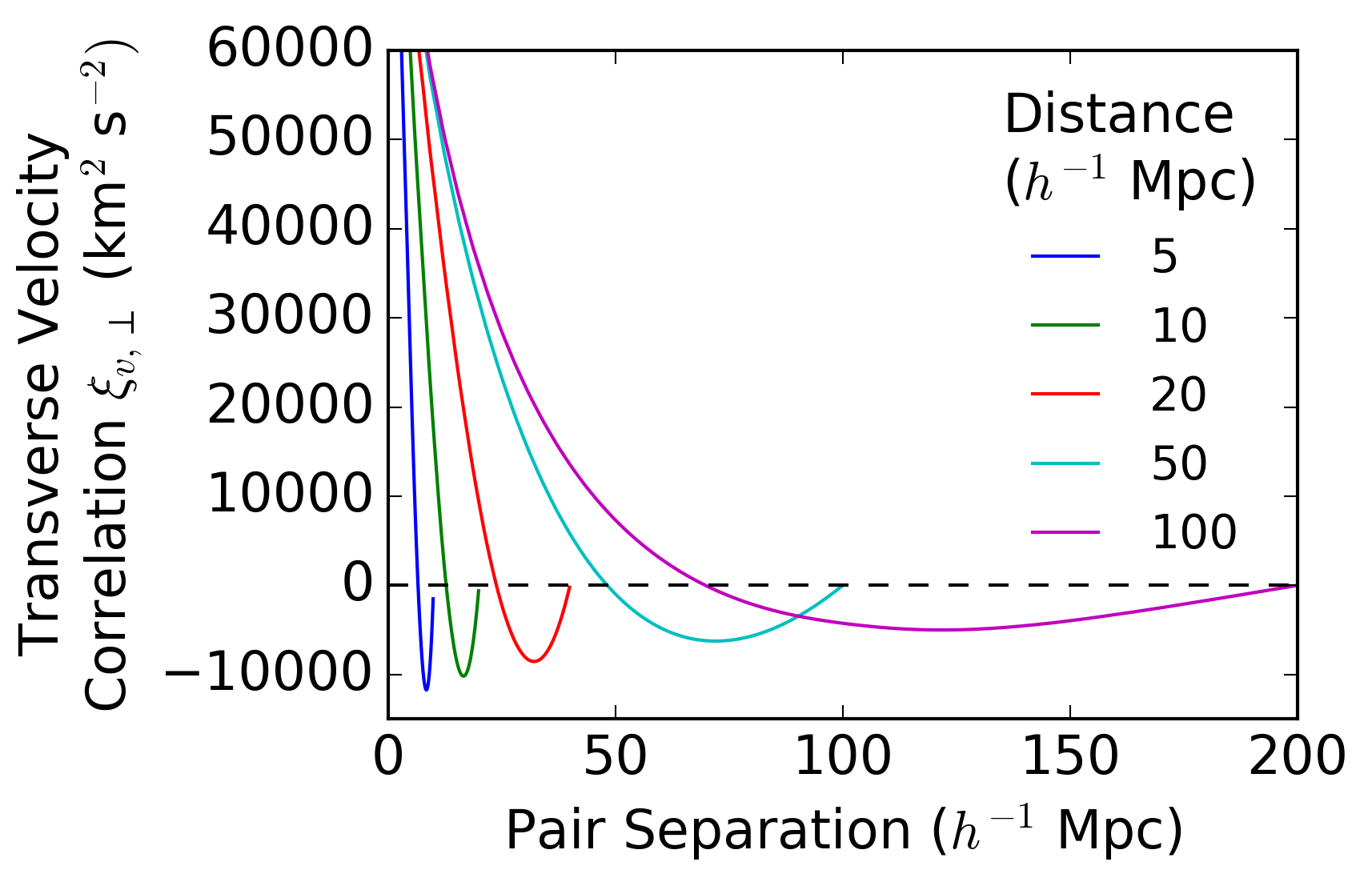}
\includegraphics[width=0.55\textwidth,trim=0 0 0 0]{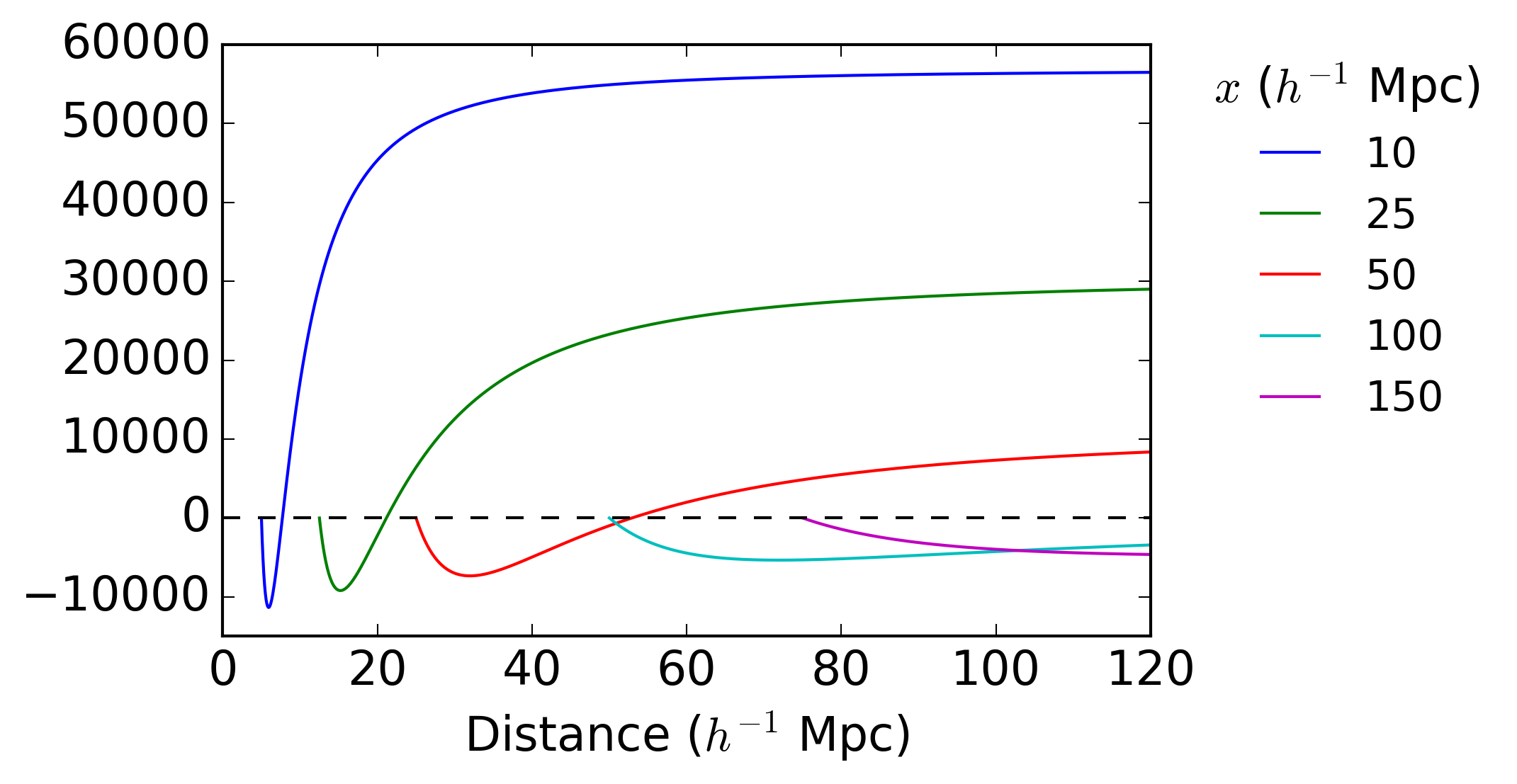}
\caption{Transverse velocity correlation $\xi_{v,\perp}(\vec{x}_1,\vec{x}_2)$ versus physical separation $x$ (left) and versus 
object distance (right) for pairs of objects at equal distance ($|\vec{x}_1| = |\vec{x}_2|$).
}\label{fig:xi_equal_D}
\end{figure}
 
It is instructive to examine the above results for equidistant pairs of objects, $|\vec{x}_1| = |\vec{x}_2|$, 
which implies that $\theta_1 = \pi - \theta_2$ and $\Delta \theta = \theta_1 - \theta_2 = \pi - 2\theta_2 = 2\theta_1 - \pi$.
In this special case, Equation \ref{eqn:result} simplifies to 
\begin{equation}
  \xi_{v,\perp}(\vec{x}_1,\vec{x}_2)\Bigr\rvert_{|\vec{x}_1| = |\vec{x}_2|}
      = \cos^4{\Delta\theta\over 2}\ \xi_{v,(i)}(x) 
              - \frac{1}{4}\sin^2\Delta\theta \ \xi_{v,(iv)}(x). \label{eqn:result_equal}
\end{equation}
Figure \ref{fig:xi_equal} shows this correlation function versus pair physical separation and versus pair angular separation.
Pairs of objects with smaller physical separations, $x \lesssim 50$~Mpc, and smaller angular separations, 
$\Delta\theta \lesssim 60^\circ$, show the largest transverse velocity correlation.  The positive value of the correlation in these
cases indicates co-streaming motions induced by density inhomogeneities (negative values would indicate converging or 
diverging motions, as is seen at low amplitude for large angular separations in Figure \ref{fig:xi_equal}.  
 
Equation \ref{eqn:result_equal} can be rewritten in terms of the ratio of the physical separation of pairs $x$ to the radial
distance to each object ($x_1 = |\vec x_1| = |\vec x_2|$):
\begin{equation}
  \xi_{v,\perp}(\vec{x}_1,\vec{x}_2)\Bigr\rvert_{|\vec{x}_1| = |\vec{x}_2|}
      = \left[1-\left({x\over2x_1}\right)^2\right]^2 \xi_{v,(i)}(x) 
              -  \left[1-\left({x\over2x_1}\right)^2\right]  \left({x\over2x_1}\right)^2 \xi_{v,(iv)}(x). \label{eqn:result_equal_D}
\end{equation}
When the two objects in a pair are equidistant, the largest possible pair separation is twice the distance to each object 
($\Delta\theta = \pi$), and the smallest possible distance is half of the pair separation.  When these extremal conditions are 
met, the projection of the transverse velocities onto $\hat x$ is zero and the correlation is null.  On the other hand, when 
pairs have separations that are small compared to their distance, $\Delta\theta$ is small and 
the correlation asymptotes to $\xi_{v,(i)}(x)$.  Figure \ref{fig:xi_equal_D} (right) demonstrates that for $x_1 \gtrsim 4 x$, 
the correlation becomes nearly constant at its largest amplitude.

\subsection{Randomly Oriented Pairs}\label{subsec:random}

Equidistant pairs represent a special case of the physical reality of randomly oriented pairs.  For fixed separation, we can 
examine the ensemble average of randomly oriented pair separation vector $\hat x$ using Equation \ref{eqn:result}.
In this case, $\theta_1$ and $\theta_2$ are not independent random variables; they are determined by the orientation of
the pair axis and the ratio of the pair separation $x$ to the distance to the pair midpoint $D$:
\begin{eqnarray}
  \sin\theta_1 &=& {\sin\phi \over \sqrt{ 1+{x\over D}\cos\phi+{1\over4} \left(x\over D\right)^2}} \label{eqn:sin_theta1}\\
  \sin\theta_2 &=& {\sin\phi \over \sqrt{ 1-{x\over D}\cos\phi+{1\over4} \left(x\over D\right)^2}}. \label{eqn:sin_theta2}
\end{eqnarray}
The orientation of the pair axis is determined by the angle $\phi$, which we choose to be zero when $\hat x$ is radial
(along the line of sight) and $\vec x_1 > \vec x_2$ (see Figure \ref{fig:diagram}).  Random pair orientation corresponds to a uniform distribution in
$\phi\ \epsilon\  [0,\pi]$.

\begin{figure}
\includegraphics[width=0.5\textwidth]{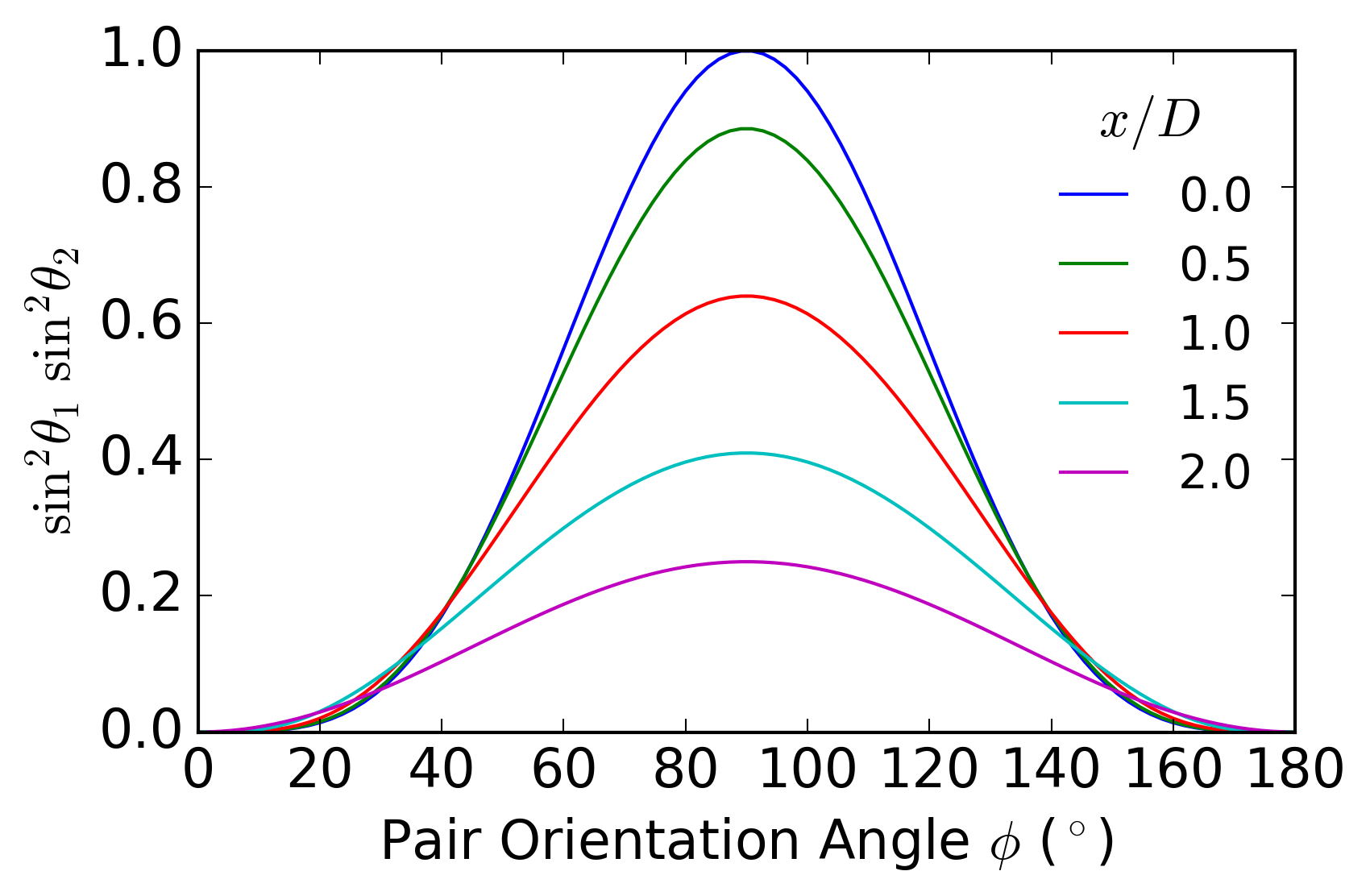}
\includegraphics[width=0.5\textwidth,trim=0 0 0 0]{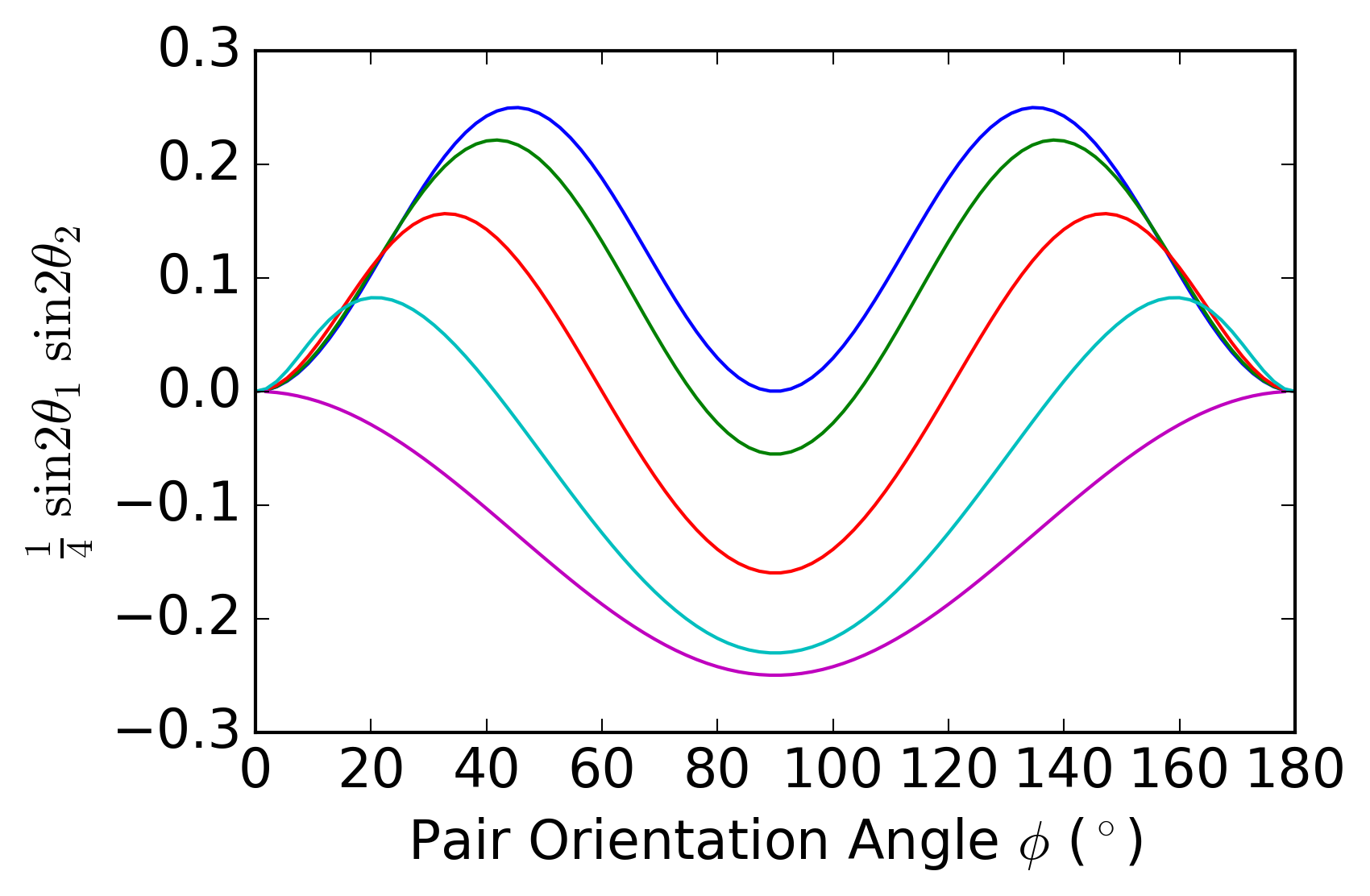}
\caption{Angular terms in Equation \ref{eqn:result} versus pair orientation angle $\phi$, plotted for a range of 
pair separation-distance ratios.  The left panel shows the angular modulation on the peculiar velocity correlation 
along the pair axis, $\xi_{v,(i)}$, and the right panel shows the angular modulation on the term perpendicular to the 
pair axis, $\xi_{v,(iv)}$.
}\label{fig:random_orientation}
\end{figure}

Using Equations \ref{eqn:sin_theta1} and \ref{eqn:sin_theta2}, the angular terms in Equation \ref{eqn:result} can be
expressed in therms of $\phi$ and $x/D$:
\begin{eqnarray}
  \sin^2\theta_1 \sin^2\theta_2 &=& {\sin^4\phi \over 1+\left({x\over D}\right)^2\left(\sin^2\phi-{1\over2}\right)+{1\over16}\left(x\over D\right)^4} \label{eqn:sinsq_x2}\\
  {1\over4}\sin2\theta_1 \sin2\theta_2 &=& {\left(1-{1\over4}\left(x\over D\right)^2\right)\sin^2\phi-\sin^4\phi \over 1+\left({x\over D}\right)^2\left(\sin^2\phi-{1\over2}\right)+{1\over16}\left(x\over D\right)^4}. \label{eqn:sin_2x2}
\end{eqnarray}
These expressions weight the contributions of the parallel, $\xi_{v,(i)}$, and perpendicular, $\xi_{v,(iv)}$, correlation terms 
to the total correlation $\xi_{v,\perp}$.  Figure \ref{fig:random_orientation} plots these angular terms for various $x/D$ and
shows that the ``parallel'' term (left) favors pairs oriented perpendicular to the line of sight (proper motions are along 
$\hat x$ and $\phi=\pi/2$) while the ``perpendicular'' term (right) gives positive weight to pairs oriented roughly 
$\pm45^\circ$ with respect to the line of sight.  

\begin{figure}
\plotone{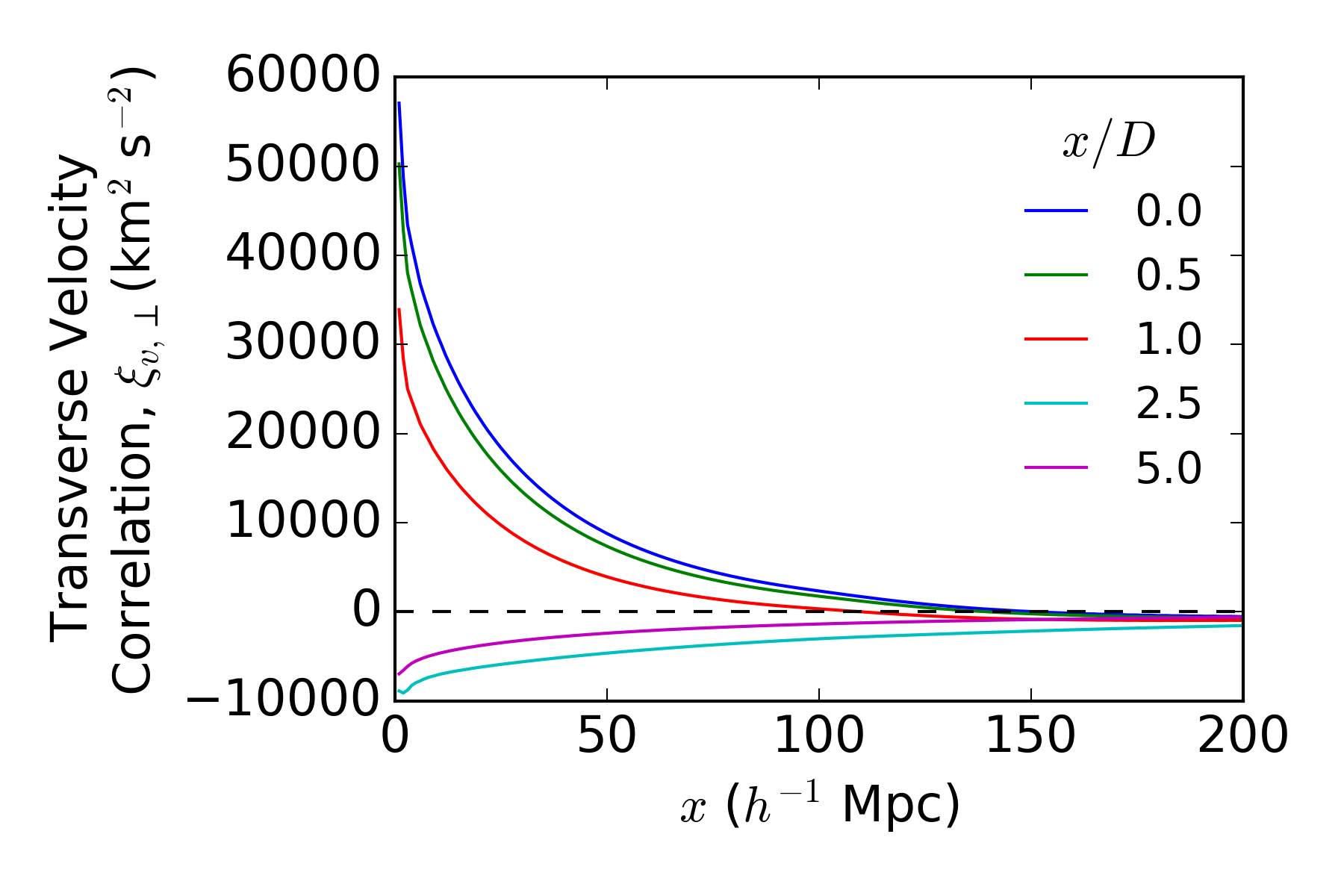}
\caption{Transverse peculiar velocity correlation statistic $\xi_{v,\perp}$ averaged over random galaxy pair 
orientations versus pair separation $x$.  Various separation-distance ratios are plotted.  For most 
large galaxy pair surveys, $x/D$ will be small and $\xi_{v,\perp}$ will be insensitive to distance.  
}\label{fig:xi_avg_orientation}
\end{figure}
 
For orientation angle $\phi$ uniformly distributed between 0 and $\pi$, we integrate Equations \ref{eqn:sinsq_x2} 
and \ref{eqn:sin_2x2} to obtain 
mean values for these angular weighting terms as a function of $x/D$.  The result predicts the outcome of observations 
of randomly oriented pairs binned in separation and/or distance.  
For all but very nearby or large-separation pairs, the orientation-averaged angular terms are insensitive to $x/D$.  For example, the perpendicular angular term's mean value differs by only $\sim1\%$ between 
$x/D = 0.2$ and $x/D\rightarrow0$.  This range will encompass most pairs in a large survey sample.

Figure \ref{fig:xi_avg_orientation} shows the total transverse peculiar velocity correlation as a function of pair separation 
for a range of $x/D$ values after averaging over randomly oriented pairs (while keeping pair separation $x$ fixed).  This 
is what one expects to observe using proper motions to calculate the transverse peculiar velocity correlation statistic 
(Equation \ref{eqn:statistic}).  The random orientation of pairs dilutes this signal compared to the equidistant 
pairs case, as expected (Figures \ref{fig:xi_equal} and \ref{fig:xi_equal_D}), but the correlation function is fairly insensitive to 
$x/D$ except for very large-separation or nearby pairs (both of which have large angular separations).  See Section
\ref{sec:alt_stat} for a parallel treatment of the alternate statistic $\xi^\prime_{v,\perp}$ that is substantially less
dependent on $x/D$ and shows a larger amplitude.

\section{An Alternate Statistic}\label{sec:alt_stat}
We define an alternate two-point correlation statistic $\xi^\prime_{v,\perp}$ based on the direct (unprojected) 
inner product of pairs of transverse velocities:
\begin{equation}  \label{eqn:altstatistic} 
  \xi^\prime_{v,\perp}(\vec{x}_1,\vec{x}_2) \equiv \langle \vec{v}_\perp(\vec{x}_1) \cdot \vec{v}_\perp(\vec{x}_2) \rangle.
\end{equation}
This statistic will produce negative values for pairs of objects that are converging or diverging
and positive values for co-streaming motions.  Using Equation \ref{eqn:vperp}, this becomes
\begin{equation}
  \xi^\prime_{v,\perp}(\vec{x}_1,\vec{x}_2) = \langle \left[\vec{v}(\vec x_1) - (\vec{v}(\vec x_1)\cdot\hat{x}_1) \hat{x}_1\right]\cdot
                                                                \left[\vec{v}(\vec x_2) - (\vec{v}(\vec x_2)\cdot\hat{x}_2)\hat{x}_2 \right]\rangle.
\end{equation}
This expression expands into four parts that can be treated separately:
\begin{eqnarray}\label{eqn:altallparts}
  \xi^\prime_{v,\perp}(\vec{x}_1,\vec{x}_2) = \langle \overbrace{\vec{v}(\vec x_1)\cdot\vec{v}(\vec x_2)}^{(i)} \rangle 
                                        + \langle \overbrace{-(\vec{v}(\vec x_1)\cdot\hat{x}_1)(\vec{v}(\vec x_2)\cdot\hat{x}_1)}^{(ii)} \rangle 
                                        + \langle \overbrace{-(\vec{v}(\vec x_2)\cdot\hat{x}_2)(\vec{v}(\vec x_1)\cdot\hat{x}_2)}^{(iii)} \rangle \nonumber \\
                                        + \langle \overbrace{(\vec{v}(\vec x_1)\cdot\hat{x}_1)(\vec{v}(\vec x_2)\cdot\hat{x}_2)\cos\Delta\theta}^{(iv)}
   \rangle
\end{eqnarray}
where $\cos\Delta\theta = \hat x_1 \cdot \hat x_2$.  
Following the derivation in Section \ref{subsec:derivation}, we work with the velocities in wavenumber space and relate velocity 
correlations to the matter power spectrum.  For parts $(ii)$--$(iv)$, it is straightforward to show that these can be written in 
terms of the correlation integrals parallel and perpendicular to $\hat x$, $\xi_{v,(i)}(x)$ and $\xi_{v,(iv)}(x)$, 
derived in Section \ref{subsec:derivation}:
\begin{eqnarray}
  (ii)&:&\ \   \langle -(\vec{v}(\vec x_1)\cdot\hat{x}_1)(\vec{v}(\vec x_2)\cdot\hat{x}_1) \rangle = -\sin^2\theta_1\ \xi_{v,(iv)}(x) - \cos^2\theta_1\ \xi_{v,(i)}(x)\\
  (iii)&:&\ \ \langle -(\vec{v}(\vec x_2)\cdot\hat{x}_2)(\vec{v}(\vec x_1)\cdot\hat{x}_2) \rangle= -\sin^2\theta_2\ \xi_{v,(iv)}(x) - \cos^2\theta_2\ \xi_{v,(i)}(x)\\
  (iv)&:&\ \ \langle (\vec{v}(\vec x_1)\cdot\hat{x}_1)(\vec{v}(\vec x_2)\cdot\hat{x}_2)\cos\Delta\theta \rangle = \cos\Delta\theta \sin\theta_1 \sin\theta_2\ \xi_{v,(iv)}(x) + \cos\Delta\theta \cos\theta_1 \cos\theta_2\ \xi_{v,(i)}(x).
\end{eqnarray}
Part $(i)$, however, is slightly different:
\begin{eqnarray}
  \langle \vec{v}(\vec x_1) \cdot \vec{v}(\vec x_2) \rangle 
      &=& \int {d^3k\over(2\pi)^3}\ e^{i \vec{k}\cdot\vec{x}_1} 
         \int {d^3k^\prime\over(2\pi)^3}\ e^{-i \vec{k}^\prime\cdot\vec{x}_2}\ 
         \langle \vec{v}(\vec{k}) \cdot \vec{v}^{\,*}(\vec{k}^\prime) \rangle \\
      &=& f^2 H_0^2\ \int {d^3k\over(2\pi)^3}\ e^{i \vec{k}\cdot\vec{x}_1} 
         \int {d^3k^\prime\over(2\pi)^3}\ e^{-i \vec{k}^\prime\cdot\vec{x}_2}\ 
         \langle \delta(\vec{k}) \delta^*(\vec{k}^\prime) \rangle {(\vec{k} \cdot \vec{k}^\prime)\over 
               k^2 k^{\prime\,2}}\\
      &=& f^2 H_0^2\ \int_0^\infty {dk\, k^2 \over(2\pi)^3}\ P(k) 
         \int d\Omega_k\ e^{i \vec{k}\cdot\vec{x}}\  {(\vec{k} \cdot \vec{k})\over k^4} \\
      &=& f^2 H_0^2\ \int_0^\infty {dk\,\over 2\pi^2k}\ P(k)\ k\ j_0(kx).
\end{eqnarray}
To recast this result in terms of the previous integration kernels, we can use the identity
\begin{equation}
  j_0(kx) = -j^{\prime\prime}_0(kx) - 2\, {j^\prime_0(kx)\over kx}
\end{equation}
and obtain a result for part $(i)$:
\begin{equation}
  (i):\ \ \langle \vec{v}(\vec x_1)\cdot\vec{v}(\vec x_2) \rangle = 2\ \xi_{v,(iv)}(x) +  \xi_{v,(i)}(x).
\end{equation}

Summing all parts of Equation \ref{eqn:altallparts}, we have
\begin{equation}
  \xi^\prime_{v,\perp}(\vec{x}_1,\vec{x}_2) 
      = \sin\theta_1\sin\theta_2\cos\Delta\theta\ \xi_{v,(i)}(x) 
              +  (1+\cos\theta_1\cos\theta_2\cos\Delta\theta)\ \xi_{v,(iv)}(x) \label{eqn:altresult}
\end{equation}
or
\begin{eqnarray}
  \xi^\prime_{v,\perp}(\vec{x}_1,\vec{x}_2) 
      = - f^2 H_0^2 \left[\sin\theta_1\sin\theta_2\cos\Delta\theta\ \int_0^\infty {dk\over 2\pi^2k}\ P(k)\ k\ j_0^{\prime\prime}(kx)\right. \nonumber \\
              +\left.(1+\cos\theta_1\cos\theta_2\cos\Delta\theta) \ \int_0^\infty {dk\over 2\pi^2k}\ P(k)\ {j_0^\prime(kx)\over x}\right].\label{eqn:altresult_ints}
\end{eqnarray}

\subsection{Equidistant Pairs}

\begin{figure}
\includegraphics[width=0.46\textwidth]{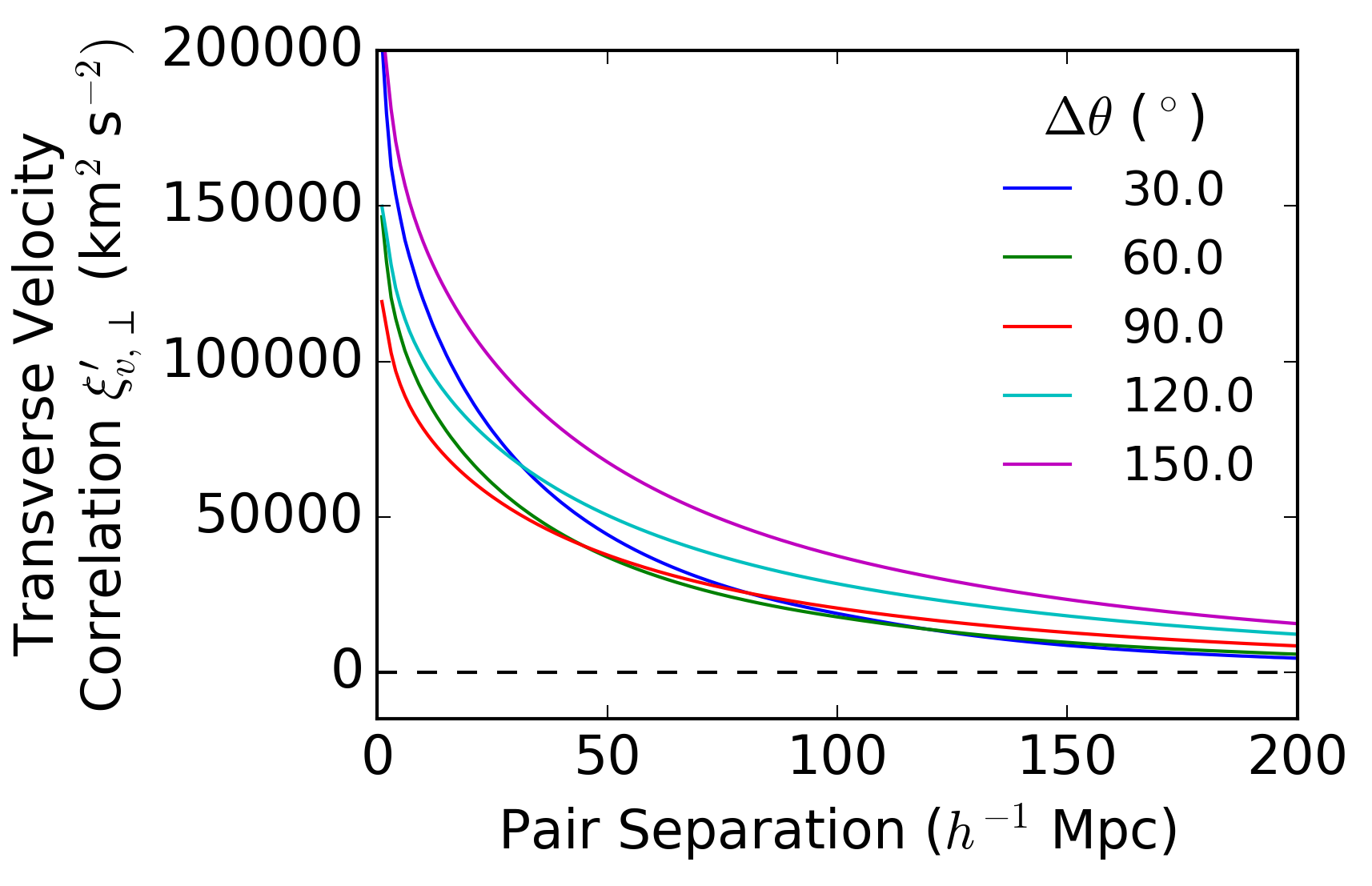}
\includegraphics[width=0.535\textwidth,trim=0 -5 0 0]{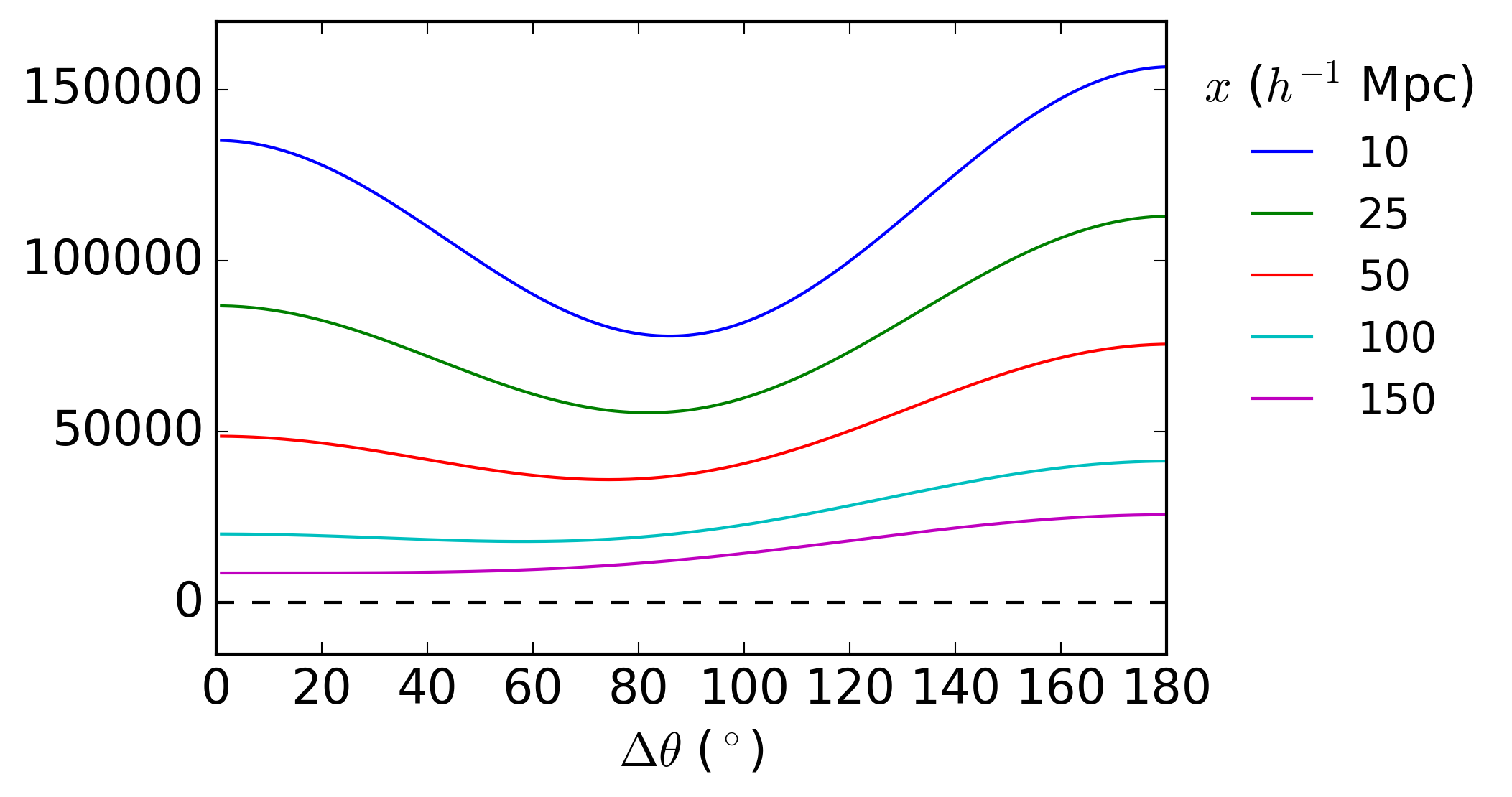}
\caption{Alternate transverse velocity correlation $\xi^\prime_{v,\perp}(\vec x_1,\vec x_2)$ versus physical separation $x$ 
(left) and versus angular separation (right) for pairs of objects at equal distance  ($|\vec{x}_1| = |\vec{x}_2|$).
}\label{fig:xi_equal_alt}
\end{figure}
 
\begin{figure}
\includegraphics[width=0.44\textwidth]{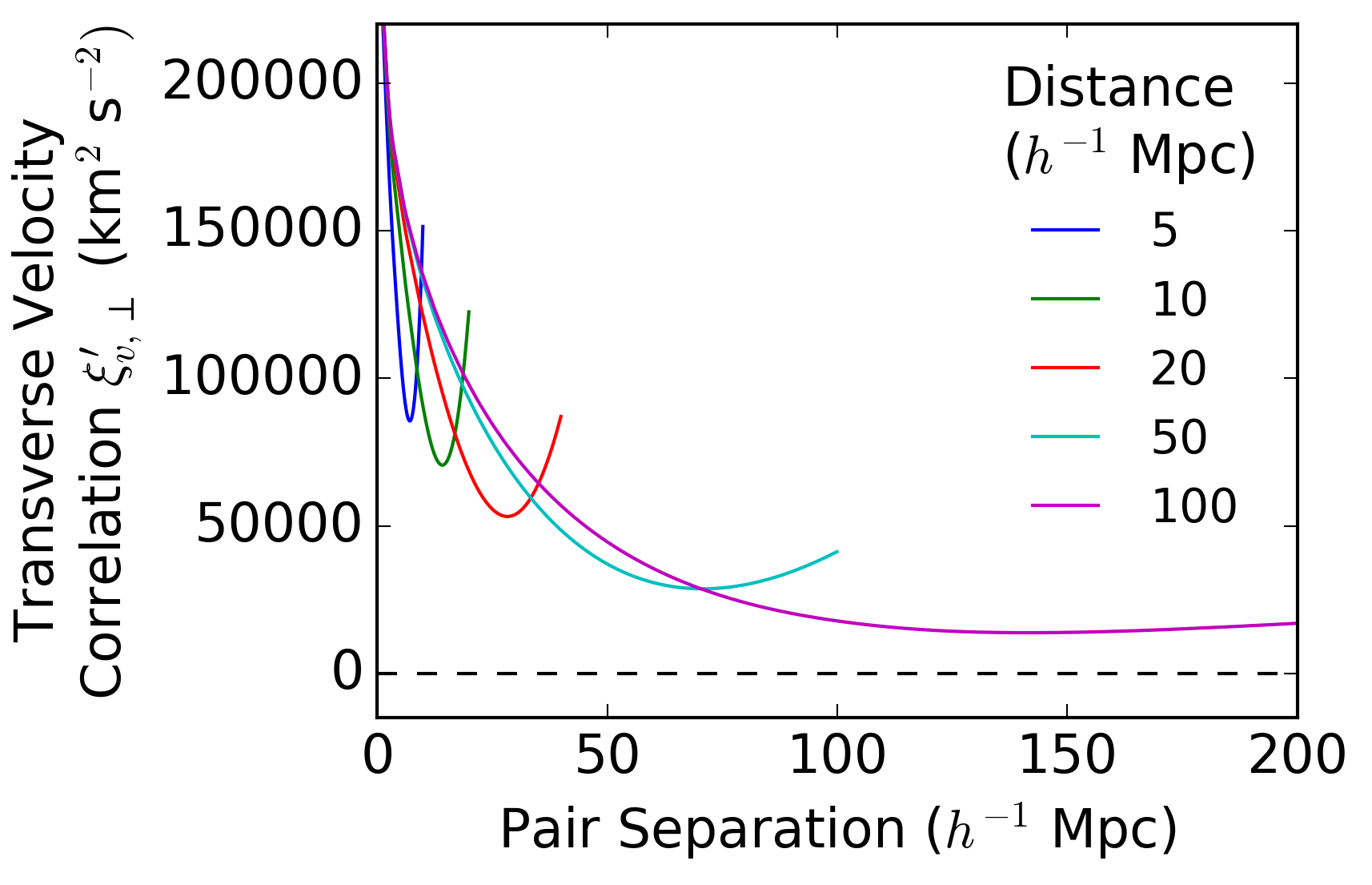}
\includegraphics[width=0.55\textwidth,trim=0 0 0 0]{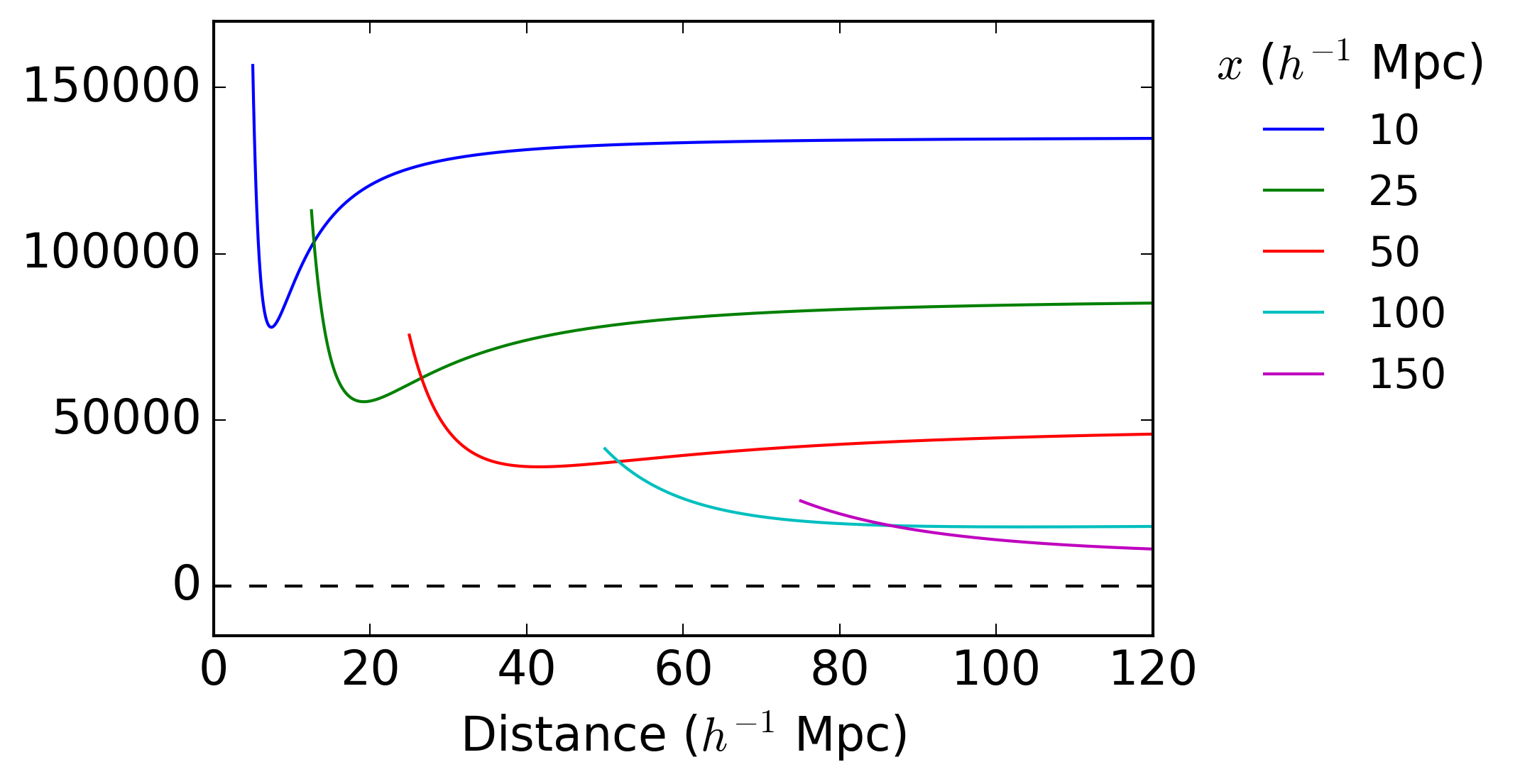}
\caption{Alternate transverse velocity correlation $\xi^\prime_{v,\perp}(\vec{x}_1,\vec{x}_2)$ versus physical 
separation $x$ (left) and versus object distance (right) for pairs of objects at equal distance ($|\vec{x}_1| = |\vec{x}_2|$).
}\label{fig:xi_equal_D_alt}
\end{figure}
 
In the simplified case where the two objects lie at the same distance, Equation \ref{eqn:altresult} becomes
\begin{equation}
  \xi^\prime_{v,\perp}(\vec{x}_1,\vec{x}_2) \Bigr\rvert_{|\vec{x}_1| = |\vec{x}_2|}
      = \cos^2{\Delta\theta\over2} \cos\Delta\theta\ \xi_{v,(i)}(x) 
              +  \left(1-\sin^2{\Delta\theta\over2}\cos\Delta\theta\right)\ \xi_{v,(iv)}(x). \label{eqn:altresult_equal}
\end{equation}
Figure \ref{fig:xi_equal_alt} shows the correlation versus pair separation (left) and angular separation (right) in 
this case.  This alternate correlation statistic is remarkably different from the statistic plotted in Figure \ref{fig:xi_equal}:
it is never negative, it is fairly insensitive to angular separation, and $\xi^\prime_{v,\perp}$ shows a larger amplitude 
than $\xi_{v,\perp}$ for all $x$ and $\Delta\theta$.  

Following the treatment in Section \ref{subsec:equidistant}, Equation \ref{eqn:altresult_equal} 
can be rewritten in terms of the ratio of the physical separation of pairs $x$ to the 
distance to each object ($x_1 = |\vec x_1| = |\vec x_2|$):
\begin{equation}
  \xi^\prime_{v,\perp} (\vec{x}_1,\vec{x}_2) \Bigr\rvert_{|\vec{x}_1| = |\vec{x}_2|}
      = \left[1-3\left({x\over2x_1}\right)^2+2\left({x\over2x_1}\right)^4\right] \xi_{v,(i)}(x) 
              +  \left[1-\left({x\over2x_1}\right)^2+2\left({x\over2x_1}\right)^4\right] \xi_{v,(iv)}(x). \label{eqn:altresult_equal_D}
\end{equation}
When the two objects in a pair are equidistant, the largest possible pair separation is twice the distance to each object 
($\Delta\theta = \pi$), and the smallest possible distance to each object is half of the pair separation.  
When these extremal conditions are met, $\xi^\prime_{v,\perp} = 2\ \xi_{v,(iv)}$.  
On the other hand, when pairs have separations that are small compared to their distance, $\Delta\theta$ is small and 
the correlation asymptotes to $\xi_{v,(i)}+\xi_{v,(iv)} < 2\ \xi_{v,(iv)}$.  Figure \ref{fig:xi_equal_D_alt} (right) demonstrates that for $x_1 \gtrsim 4 x$, the correlation becomes nearly constant (but is not a maximum). 
The asymptote for $x_1 \gtrsim 4 x$ has a lower amplitude than the maximal value when $x_1 = x/2$.

\begin{figure}
\includegraphics[width=0.5\textwidth]{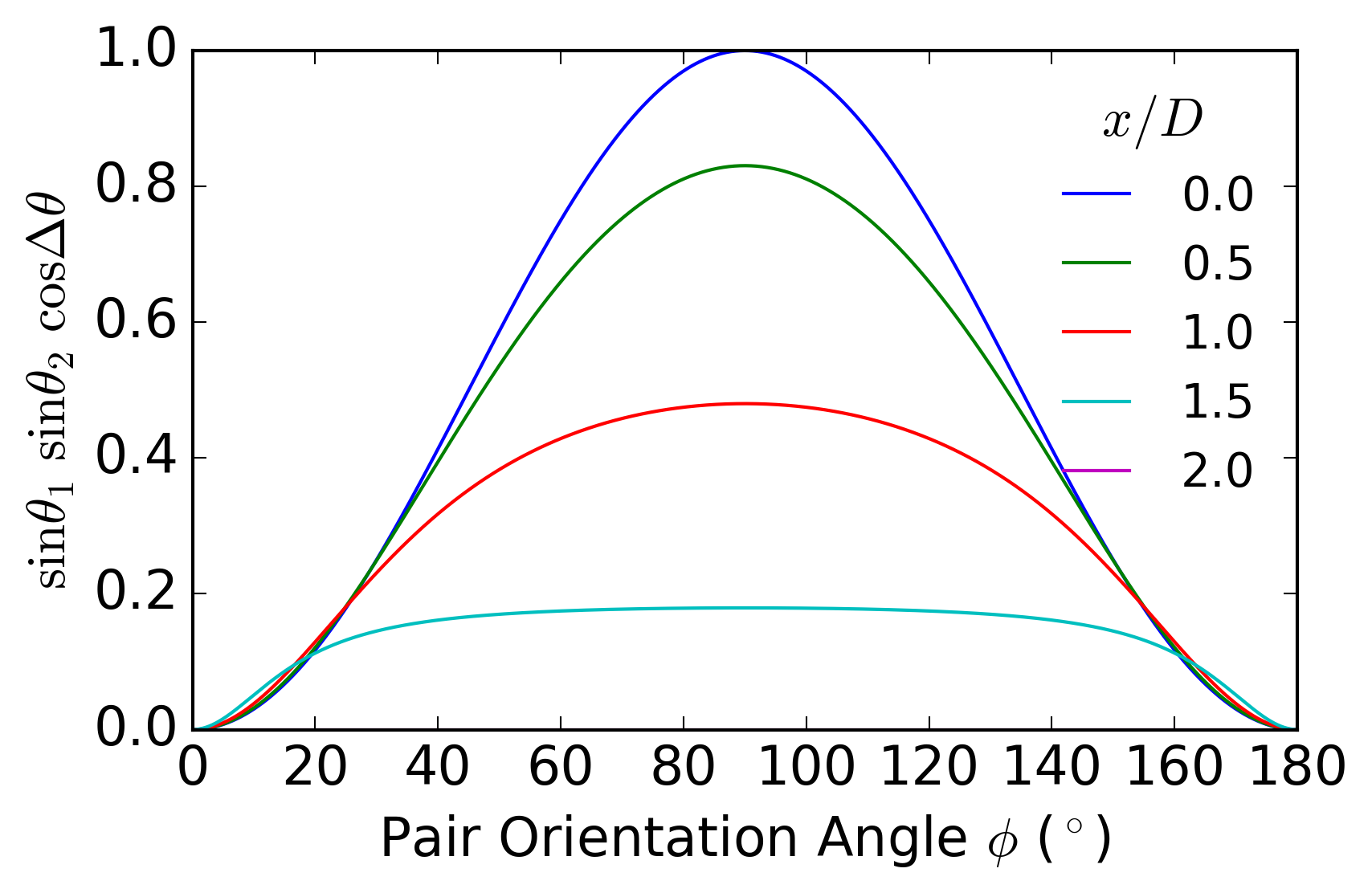}
\includegraphics[width=0.5\textwidth,trim=0 0 0 0]{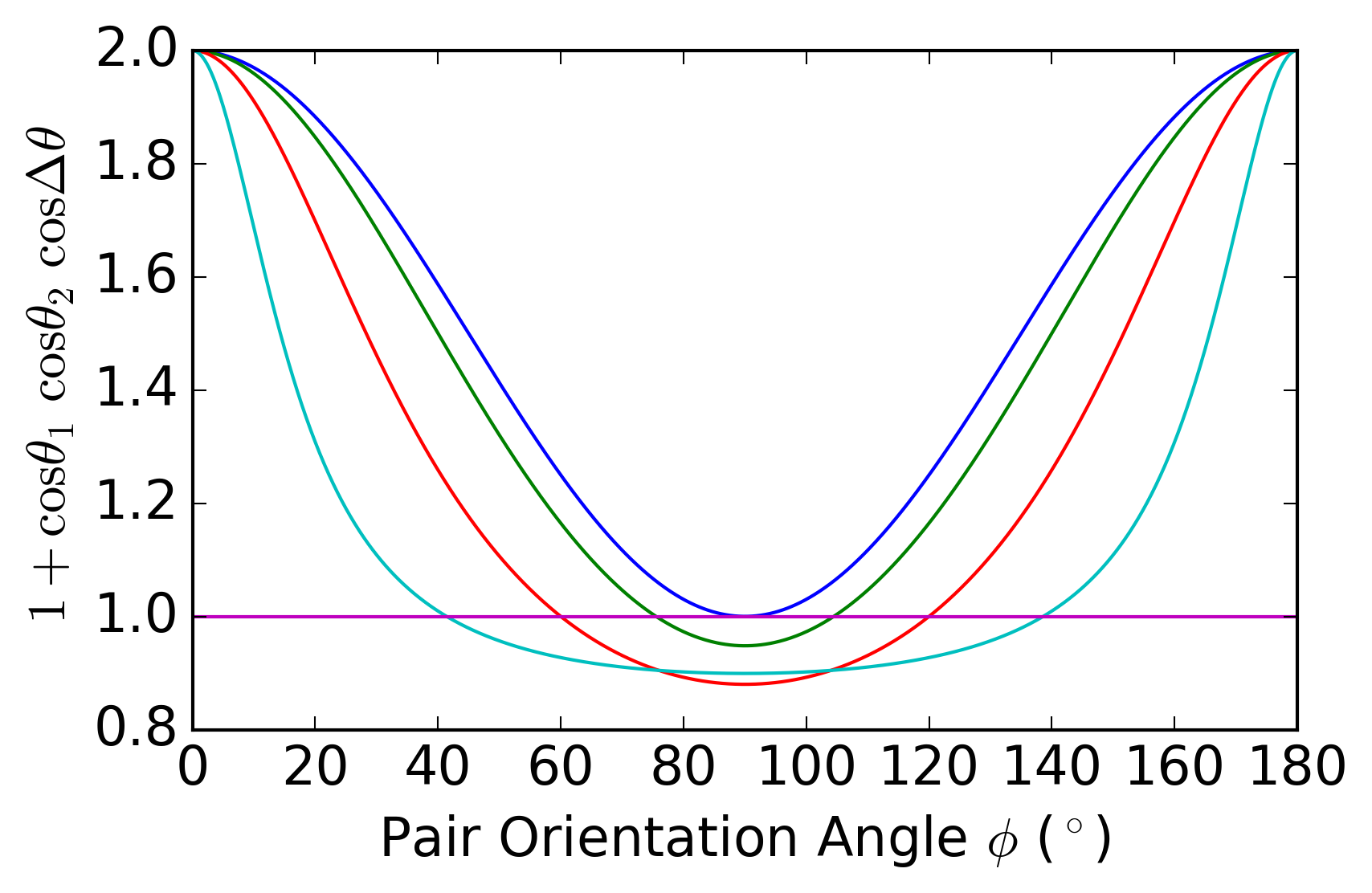}
\caption{Angular terms in Equation \ref{eqn:altresult} versus pair orientation angle $\phi$, plotted for a range of 
pair separation-distance ratios.  The left panel shows the angular modulation on the peculiar velocity correlation 
along the pair axis, $\xi_{v,(i)}$, and the right panel shows the angular modulation on the term perpendicular to the 
pair axis, $\xi_{v,(iv)}$.
}\label{fig:random_orientation_alt}
\end{figure}

\begin{figure}
\epsscale{0.875}
\plotone{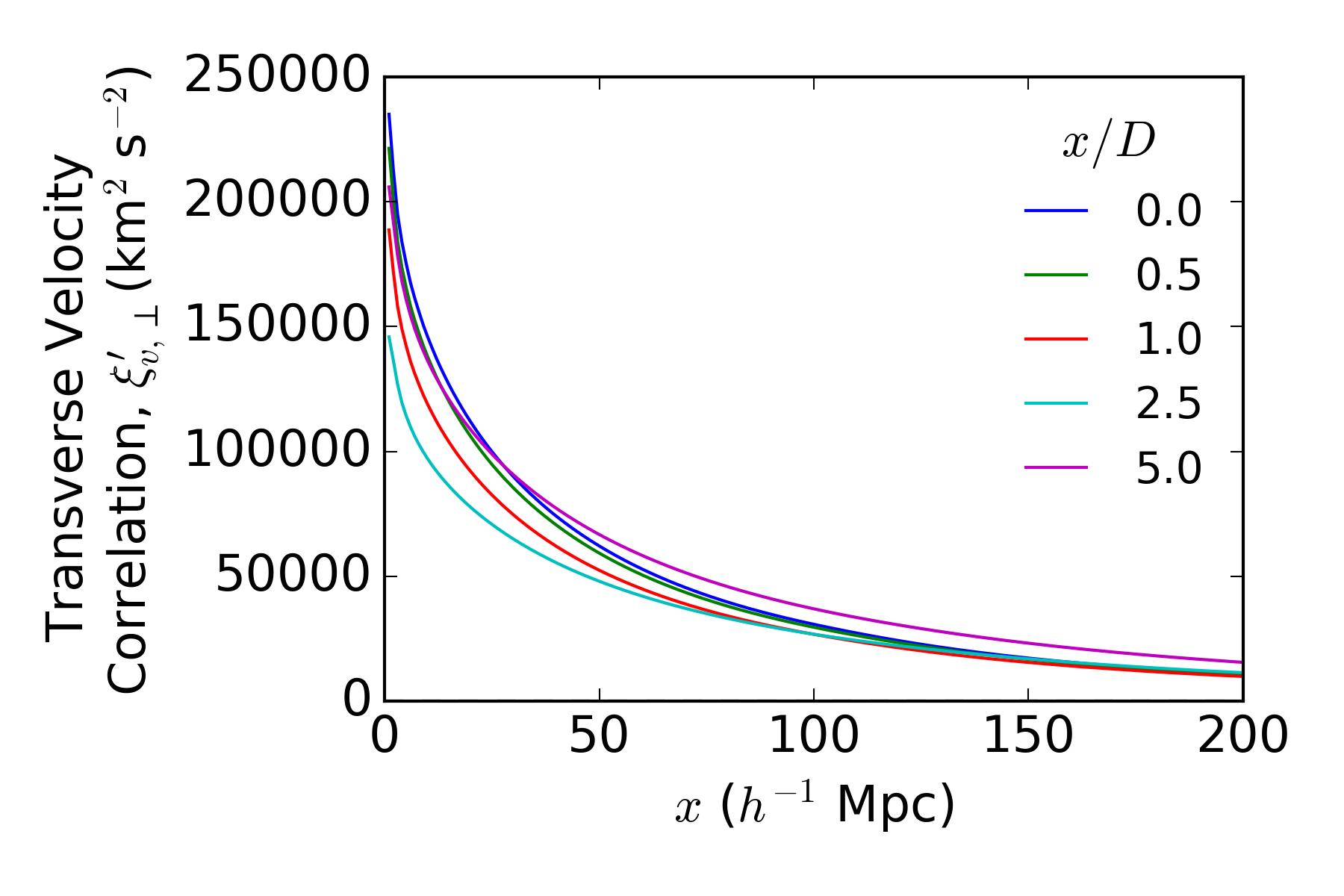}
\caption{Alternate transverse peculiar velocity correlation statistic $\xi^\prime_{v,\perp}$ averaged over random galaxy pair 
orientations versus pair separation $x$.  Various separation-distance ratios are plotted, showing that this correlation 
is insensitive to the ratio and is primarily a function of pair separation.
}\label{fig:xi_avg_orientation_alt}
\end{figure}
 
\clearpage

\subsection{Randomly Oriented Pairs}

The angular terms in Equation \ref{eqn:altresult} can be expressed in terms of the pair orientation angle $\phi$ and the 
separation-to-distance ratio $x/D$:
\begin{eqnarray}
  \sin\theta_1 \sin\theta_2 \cos\Delta\theta &=& {\left(1-{1\over4}\left(x\over D\right)^2\right)\sin^2\phi \over 1+\left({x\over D}\right)^2\left(\sin^2\phi-{1\over2}\right)+{1\over16}\left(x\over D\right)^4} \label{eqn:sin_alt}\\
  1+\cos\theta_1 \cos\theta_2 \cos\Delta\theta &=& 2 - {\left(1+{3\over4}\left(x\over D\right)^2\right)\sin^2\phi \over 1+\left({x\over D}\right)^2\left(\sin^2\phi-{1\over2}\right)+{1\over16}\left(x\over D\right)^4}. \label{eqn:cos_alt}
\end{eqnarray}
Figure \ref{fig:random_orientation_alt} 
plots the angular terms versus pair orientation, showing that even for a radially oriented pair ($\phi = 0$) 
this statistic is sensitive to correlated peculiar velocity.  
For uniformly random orientations, we calculate the mean values for these angular terms and sum the components of Equation
\ref{eqn:altresult} to obtain the transverse velocity correlation shown in Figure \ref{fig:xi_avg_orientation_alt}.  
This alternate statistic is roughly four times larger than the projected statistic and it is insensitive to $x/D$.  This 
statistic is observation-friendly because one can simply average pair measurements binned by pair separation.

\section{Discussion }\label{sec:discussion}

Given the results above, 
proper motion observations should focus on the smallest physical separation pairs of galaxies or AGN that also 
have small angular separations (large distances), or small $x/D$.  This strategy will maximize the expected correlation signal.  
The caveat to this guidance, which suggests that the best tracers of transverse peculiar motions
would be distant small-separation pairs of objects, is that the observed quantity is the proper motion.  The rest-frame
transverse velocity is proportional to the proper motion and the distance,
\begin{equation} \label{eqn:v_mu}
  \vec v_\perp = D_M\ \vec \mu, 
\end{equation}
where $D_M$ is the proper motion distance (equivalent to the comoving distance in a flat cosmology and related to 
the angular diameter distance as $D_M = D_A (1+z)$ \citep{hogg1999}).  Uncertainties in proper motion measurements
will therefore scale linearly with distance when translated into transverse velocity uncertainties.  
A typical peculiar velocity of 300~km~s$^{-1}$ equates to a proper motion of 6.3~$\mu$as~yr$^{-1}$ at 10 Mpc.  
A sample of objects with proper motion uncertainty $\sigma_\mu = 100$~$\mu$as~yr$^{-1}$ would require 
roughly 250 independent measurements to reach this proper motion value and roughly 2300 measurements to achieve 3$\sigma$ significance.
Such averaging to reduce statistical errors would only be possible if one knew {\it a priori} the orientation of each 
proper motion vector in the sample.  Since this is not known, it is appropriate to examine correlation of extragalactic 
pairs.  

Since the number of galaxies in a volume scales as $N\propto D^3$, and the number of possible pairs 
scales as $N_p \sim N^2$, then the number of pairs scales as $N_p \propto D^6$.  
The uncertainty in the transverse velocity correlation scales as $\sigma_{\xi_\perp} \propto v_\perp\,\sigma_{v_\perp} N_p^{-1/2}$.
Using Equation \ref{eqn:v_mu}, and assuming that the fractional proper motion uncertainty is much larger than the distance
uncertainty, $\sigma_{v_\perp} = \sigma_\mu D$.  The uncertainty in the correlation therefore scales roughly as 
$\sigma_{\xi_\perp} \propto \mu\,\sigma_\mu D^{-1}$.  This relationship assumes that $x/D \lesssim 1$, that pairs 
represent independent measurements of the peculiar velocity field (not strictly true), that the proper motion measurements
in value and uncertainty are uniform among the sample, and it neglects the random orientations of pairs.  Including these
effects will impact the constant of proportionality and may diminish the favorable impact of distance on the uncertainty in 
the measured correlation statistic.  

The optimal strategy for measuring the transverse peculiar velocity created by large scale structure will depend on the 
data in hand:  the redshift distribution, pair separations, sample size, and proper motion precision.  VLBI proper motions 
will likely have redshifts $z\sim1$, proper motion uncertainties per object of $\sigma_\mu \sim 10$~$\mu$as~yr$^{-1}$, and 
sample size of 500--1000 objects \citep[e.g.,][]{darling2013,truebenbach2017}.  {\it Gaia} AGN proper motions will have lower
redshifts, smaller pair separations, and sample size $\sim10^5$, but larger per-object proper motion uncertainties 
of $\sim 200$~$\mu$as~yr$^{-1}$ \citep[e.g.,][]{paine2018,truebenbach2018}.

\section{Conclusions and Future Work}

We have developed two-point transverse peculiar velocity correlation statistics that connect the matter power spectrum to 
extragalactic proper motions.  We have explored the impact of pair separation, distance, and angular separation on 
these correlations and suggested some strategies for future extragalactic proper motion surveys.  Except for very 
nearby pairs of objects that subtend large angles, close pairs are more sensitive to large scale structure than 
widely-separated pairs, as one might expect.  We develop correlation statistics for randomly oriented pairs of objects 
with various separation-to-distance ratios.  Finally, we suggest that the sensitivity to the transverse velocity correlation 
improves with distance in volume-limited surveys, but the optimal detection strategy will depend on the nature of the
proper motion data in hand.

The work presented here has limitations that point to obvious directions for future work:
\begin{itemize}

\item The two-point correlation statistics presented here are scalars while the inputs are proper motion vectors, so this work 
is not taking full advantage of the available observational information.  We suggest that future work should consider 
a vector correlation measure.  What this might look like, we do not know.

\item It will also be important to extend this work to higher redshifts, which will require modifications to $f$, $H_0$, and the 
power spectrum $P(k)$.  The treatment above is correct for low redshift, but quasars in general and radio sources in 
particular are significantly redshifted, and VLBI or {\it Gaia} samples will typically have mean 
redshifts of $z\sim1$, which is much more distant than traditional radial peculiar velocity surveys.  

\item Very close pairs of objects will not obey linear perturbation theory.  A complete treatment will need to rely on numerical 
simulations.  

\end{itemize}
Ultimately, it may be possible to combine this work with radial peculiar velocities to obtain a true three-dimensional 
peculiar velocity map of the local universe.

\acknowledgments
 
The authors acknowledge support from the NSF grant AST-1411605 and the NASA grant 14-ATP14-0086.  
We thank the anonymous referee for helpful suggestions and for calculation validation.
We acknowledge the use of the Legacy Archive for Microwave Background Data Analysis (LAMBDA), part of the High Energy Astrophysics Science Archive Center (HEASARC). HEASARC/LAMBDA is a service of the Astrophysics Science Division at the NASA Goddard Space Flight Center.

\software{CAMB}

\end{document}